

%
%


\def\famname{
 \textfont0=\textrm \scriptfont0=\scriptrm
 \scriptscriptfont0=\sscriptrm
 \textfont1=\textmi \scriptfont1=\scriptmi
 \scriptscriptfont1=\sscriptmi
 \textfont2=\textsy \scriptfont2=\scriptsy \scriptscriptfont2=\sscriptsy
 \textfont3=\textex \scriptfont3=\textex \scriptscriptfont3=\textex
 \textfont4=\textbf \scriptfont4=\scriptbf \scriptscriptfont4=\sscriptbf
 \skewchar\textmi='177 \skewchar\scriptmi='177
 \skewchar\sscriptmi='177
 \skewchar\textsy='60 \skewchar\scriptsy='60
 \skewchar\sscriptsy='60
 \def\rm{\fam0 \textrm} \def\bf{\fam4 \textbf}}
\def\sca#1{scaled\magstep#1} \def\scah{scaled\magstephalf} 
\def\twelvepoint{
 \font\textrm=cmr12 \font\scriptrm=cmr8 \font\sscriptrm=cmr6
 \font\textmi=cmmi12 \font\scriptmi=cmmi8 \font\sscriptmi=cmmi6 
 \font\textsy=cmsy10 \sca1 \font\scriptsy=cmsy8
 \font\sscriptsy=cmsy6
 \font\textex=cmex10 \sca1
 \font\textbf=cmbx12 \font\scriptbf=cmbx8 \font\sscriptbf=cmbx6
 \font\it=cmti12
 \font\sectfont=cmbx12 \sca1
 \font\sectmath=cmmib10 \sca2
 \font\sectsymb=cmbsy10 \sca2
 \font\refrm=cmr10 \scah \font\refit=cmti10 \scah
 \font\refbf=cmbx10 \scah
 \def\twelverm{\textrm} \def\twelveit{\it} \def\twelvebf{\textbf}
 \famname \textrm 
 \advance\voffset by .06in \advance\hoffset by .28in
 \normalbaselineskip=17.5pt plus 1pt \baselineskip=\normalbaselineskip
 \parindent=21pt
 \setbox\strutbox=\hbox{\vrule height10.5pt depth4pt width0pt}}


\catcode`@=11

{\catcode`\'=\active \def'{{}^\bgroup\prim@s}}

\def\screwcount{\alloc@0\count\countdef\insc@unt}   
\def\screwdimen{\alloc@1\dimen\dimendef\insc@unt} 
\def\screwbox{\alloc@4\box\chardef\insc@unt}

\catcode`@=12


\overfullrule=0pt			
\vsize=9in \hsize=6in
\lineskip=0pt				
\abovedisplayskip=1.2em plus.3em minus.9em 
\belowdisplayskip=1.2em plus.3em minus.9em	
\abovedisplayshortskip=0em plus.3em	
\belowdisplayshortskip=.7em plus.3em minus.4em	
\parindent=21pt
\setbox\strutbox=\hbox{\vrule height10.5pt depth4pt width0pt}
\def\makefootline{\baselineskip=30pt \line{\the\footline}}
\footline={\ifnum\count0=1 \hfil \else\hss\twelverm\folio\hss \fi}
\pageno=1


\def\put(#1,#2)#3{\screwdimen\unit  \unit=1in
	\vbox to0pt{\kern-#2\unit\hbox{\kern#1\unit
	\vbox{#3}}\vss}\nointerlineskip}


\def\\{\hfil\break}
\def\newpage{\vfill\eject}
\def\center{\leftskip=0pt plus 1fill \rightskip=\leftskip \parindent=0pt
 \def\textindent##1{\par\hangindent21pt\footrm\noindent\hskip21pt
 \llap{##1\enspace}\ignorespaces}\par}
\def\unnarrower{\leftskip=0pt \rightskip=\leftskip}


\def\vol#1 {{\refbf#1} }		 


\def\NP #1 {{\refit Nucl. Phys.} {\refbf B{#1}} }
\def\PL #1 {{\refit Phys. Lett.} {\refbf{#1}} }
\def\PR #1 {{\refit Phys. Rev. Lett.} {\refbf{#1}} }
\def\PRD #1 {{\refit Phys. Rev.} {\refbf D{#1}} }


\hyphenation{pre-print}
\hyphenation{quan-ti-za-tion}

%
%

\def\on#1#2{{\buildrel{\mkern2.5mu#1\mkern-2.5mu}\over{#2}}}
\def\oonoo#1#2#3{\vbox{\ialign{##\crcr
	\hfil\hfil\hfil{$#3{#1}$}\hfil\crcr\noalign{\kern1pt\nointerlineskip}
	$#3{#2}$\crcr}}}
\def\oon#1#2{\mathchoice{\oonoo{#1}{#2}{\displaystyle}}
	{\oonoo{#1}{#2}{\textstyle}}{\oonoo{#1}{#2}{\scriptstyle}}
	{\oonoo{#1}{#2}{\scriptscriptstyle}}}
\def\dt#1{\oon{\hbox{\bf .}}{#1}}  
\def\ddt#1{\oon{\hbox{\bf .\kern-1pt.}}#1}    
\def\slap#1#2{\setbox0=\hbox{$#1{#2}$}
	#2\kern-\wd0{\hfuzz=1pt\hbox to\wd0{\hfil$#1{/}$\hfil}}}
\def\sla#1{\mathpalette\slap{#1}}                
\def\bop#1{\setbox0=\hbox{$#1M$}\mkern1.5mu
	\lower.02\ht0\vbox{\hrule height0pt depth.06\ht0
	\hbox{\vrule width.06\ht0 height.9\ht0 \kern.9\ht0
	\vrule width.06\ht0}\hrule height.06\ht0}\mkern1.5mu}
\def\bo{{\mathpalette\bop{}}}                        
\def~{\widetilde} 
\mathcode`\*="702A                  
\def\in{\relax\ifmmode\mathchar"3232\else{\refit in\/}\fi} 
\def\half{{\textstyle{1\over{\raise.1ex\hbox{$\scriptstyle{2}$}}}}}

\def\Gamma{\mathchar"0100}
\def\Delta{\mathchar"0101}
\def\Theta{\mathchar"0102}
\def\Lambda{\mathchar"0103}
\def\Xi{\mathchar"0104}
\def\Pi{\mathchar"0105}
\def\Sigma{\mathchar"0106}
\def\Upsilon{\mathchar"0107}
\def\Phi{\mathchar"0108}
\def\Psi{\mathchar"0109}
\def\Omega{\mathchar"010A}

\catcode128=13 \def €{\"A}                 
\catcode129=13 \def {\AA}                 
\catcode130=13 \def '{\c}           	   
\catcode131=13 \def ƒ{\'E}                   
\catcode132=13 \def "{\~N}                   
\catcode133=13 \def …{\"O}                 
\catcode134=13 \def †{\"U}                  
\catcode135=13 \def ‡{\'a}                  
\catcode136=13 \def ˆ{\`a}                   
\catcode137=13 \def ‰{\^a}                 
\catcode138=13 \def Š{\"a}                 
\catcode139=13 \def ‹{\~a}                   
\catcode140=13 \def Œ{\alpha}            
\catcode141=13 \def {\chi}                
\catcode142=13 \def Ž{\'e}                   
\catcode143=13 \def {\`e}                    
\catcode144=13 \def {\^e}                  
\catcode145=13 \def '{\"e}                
\catcode146=13 \def '{\'\i}                 
\catcode147=13 \def "{\`\i}                  
\catcode148=13 \def "{\^\i}                
\catcode149=13 \def •{\"\i}                
\catcode150=13 \def –{\~n}                  
\catcode151=13 \def —{\'o}                 
\catcode152=13 \def ˜{\`o}                  
\catcode153=13 \def ™{\^o}                
\catcode154=13 \def š{\"o}                 
\catcode155=13 \def ›{\~o}                  
\catcode156=13 \def œ{\'u}                  
\catcode157=13 \def {\`u}                  
\catcode158=13 \def ž{\^u}                
\catcode159=13 \def Ÿ{\"u}                
\catcode160=13 \def  {\tau}               
\catcode161=13 \mathchardef ¡="2203     
\catcode162=13 \def ¢{\oplus}           
\catcode163=13 \def £{\relax\ifmmode\to\else\itemize\fi} 
\catcode164=13 \def ¤{\subset}	  
\catcode165=13 \def ¥{\infty}           
\catcode166=13 \def ¦{\mp}                
\catcode167=13 \def §{\sigma}           
\catcode168=13 \def ¨{\rho}               
\catcode169=13 \def ©{\gamma}         
\catcode170=13 \def ª{\leftrightarrow} 
\catcode171=13 \def «{\relax\ifmmode\acute\else\expandafter\'\fi}
\catcode172=13 \def ¬{\relax\ifmmode\expandafter\ddt\else\expandafter\"\fi}
\catcode173=13 \def ­{\equiv}            
\catcode174=13 \def ®{\approx}          
\catcode175=13 \def ¯{\Omega}          
\catcode176=13 \def °{\otimes}          
\catcode177=13 \def ±{\ne}                 
\catcode178=13 \def ²{\le}                   
\catcode179=13 \def ³{\ge}                  
\catcode180=13 \def ´{\upsilon}          
\catcode181=13 \def µ{\mu}                
\catcode182=13 \def ¶{\delta}             
\catcode183=13 \def ·{\epsilon}          
\catcode184=13 \def ¸{\Pi}                  
\catcode185=13 \def ¹{\pi}                  
\catcode186=13 \def º{\beta}               
\catcode187=13 \def »{\partial}           
\catcode188=13 \def ¼{\nobreak\ }       
\catcode189=13 \def ½{\zeta}               
\catcode190=13 \def ¾{\sim}                 
\catcode191=13 \def ¿{\omega}           
\catcode192=13 \def À{\dt}                     
\catcode193=13 \def Á{\gets}                
\catcode194=13 \def Â{\lambda}           
\catcode195=13 \def Ã{\nu}                   
\catcode196=13 \def Ä{\phi}                  
\catcode197=13 \def Å{\xi}                     
\catcode198=13 \def Æ{\psi}                  
\catcode199=13 \def Ç{\int}                    
\catcode200=13 \def È{\oint}                 
\catcode201=13 \def É{\relax\ifmmode\cdot\else\vol\fi}    
\catcode202=13 \def Ê{\relax\ifmmode\,\else\thinspace\fi}
\catcode203=13 \def Ë{\`A}                      
\catcode204=13 \def Ì{\~A}                      
\catcode205=13 \def Í{\~O}                      
\catcode206=13 \def Î{\Theta}              
\catcode207=13 \def Ï{\theta}               
\catcode208=13 \def Ð{\relax\ifmmode\bar\else\expandafter\=\fi}
\catcode209=13 \def Ñ{\overline}             
\catcode210=13 \def Ò{\langle}               
\catcode211=13 \def Ó{\relax\ifmmode\{\else\ital\fi}      
\catcode212=13 \def Ô{\rangle}               
\catcode213=13 \def Õ{\}}                        
\catcode214=13 \def Ö{\sla}                      
\catcode215=13 \def ×{\relax\ifmmode\check\else\expandafter\v\fi}
\catcode216=13 \def Ø{\"y}                     
\catcode217=13 \def Ù{\"Y}  		    
\catcode218=13 \def Ú{\Leftarrow}       
\catcode219=13 \def Û{\Leftrightarrow}       
\catcode220=13 \def Ü{\relax\ifmmode\Rightarrow\else\sect\fi}
\catcode221=13 \def Ý{\sum}                  
\catcode222=13 \def Þ{\prod}                 
\catcode223=13 \def ß{\widehat}              
\catcode224=13 \def à{\pm}                     
\catcode225=13 \def á{\nabla}                
\catcode226=13 \def â{\quad}                 
\catcode227=13 \def ã{\in}               	
\catcode228=13 \def ä{\star}      	      
\catcode229=13 \def å{\sqrt}                   
\catcode230=13 \def æ{\^E}			
\catcode231=13 \def ç{\Upsilon}              
\catcode232=13 \def è{\"E}    	   	 
\catcode233=13 \def é{\`E}               	  
\catcode234=13 \def ê{\Sigma}                
\catcode235=13 \def ë{\Delta}                 
\catcode236=13 \def ì{\Phi}                     
\catcode237=13 \def í{\`I}        		   
\catcode238=13 \def î{\iota}        	     
\catcode239=13 \def ï{\Psi}                     
\catcode240=13 \def ð{\times}                  
\catcode241=13 \def ñ{\Lambda}             
\catcode242=13 \def ò{\cdots}                
\catcode243=13 \def ó{\^U}			
\catcode244=13 \def ô{\`U}    	              
\catcode245=13 \def õ{\bo}                       
\catcode246=13 \def ö{\relax\ifmmode\hat\else\expandafter\^\fi}
\catcode247=13 \def÷{\relax\ifmmode\tilde\else\expandafter\~\fi}
\catcode248=13 \def ø{\ll}                         
\catcode249=13 \def ù{\gg}                       
\catcode250=13 \def ú{\eta}                      
\catcode251=13 \def û{\kappa}                  
\catcode252=13 \def ü{\half}     		 
\catcode253=13 \def ý{\Gamma} 		
\catcode254=13 \def þ{\Xi}   			
\catcode255=13 \def ÿ{\relax\ifmmode{}^{\dagger}{}\else\dag\fi}


\def\ital#1Õ{{\it#1\/}}	     
\def\un#1{\relax\ifmmode\underline#1\else $\underline{\hbox{#1}}$
	\relax\fi}

\def\roonoo#1#2#3{\vbox{\ialign{##\crcr
	\hfil{$#3{#1}$}\hfil\crcr\noalign{\kern1pt\nointerlineskip}
	$#3{#2}$\crcr}}}

\def\tdt#1{\oon{\hbox{\bf .\kern-1pt.\kern-1pt.}}#1}   
\def\({\eqno(}



\def\õ#1{
	\screwcount\num
	\num=1
	\screwdimen\downsy
	\downsy=-1.5ex
	\mkern-3.5mu
	õ
	\loop
	\ifnum\num<#1
	\llap{\raise\num\downsy\hbox{$õ$}}
	\advance\num by1
	\repeat}
\def\upõ#1#2{\screwcount\numup
	\numup=#1
	\advance\numup by-1
	\screwdimen\upsy
	\upsy=.75ex
	\mkern3.5mu
	\raise\numup\upsy\hbox{$#2$}}



\newcount\marknumber	\marknumber=1
\newcount\countdp \newcount\countwd \newcount\countht 

%
%
\ifx\pdfoutput\undefined
\def\rgboo#1{}
\input epsf

\def\postscript#1{\special{" #1}}		
\postscript{
	/bd {bind def} bind def
	/fsd {findfont exch scalefont def} bd
	/sms {setfont moveto show} bd
	/ms {moveto show} bd
	/pdfmark where		
	{pop} {userdict /pdfmark /cleartomark load put} ifelse
	[ /PageMode /UseOutlines		
	/DOCVIEW pdfmark}
\def\bookmark#1#2{\postscript{		
	[ /Dest /MyDest\the\marknumber /View [ /XYZ null null null ] /DEST pdfmark
	[ /Title (#2) /Count #1 /Dest /MyDest\the\marknumber /OUT pdfmark}%
	\advance\marknumber by1}
\def\pdfklink#1#2{%
	\hskip-.25em\setbox0=\hbox{#1}%
		\countdp=\dp0 \countwd=\wd0 \countht=\ht0%
		\divide\countdp by65536 \divide\countwd by65536%
			\divide\countht by65536%
		\advance\countdp by1 \advance\countwd by1%
			\advance\countht by1%
		\def\linkdp{\the\countdp} \def\linkwd{\the\countwd}%
			\def\linkht{\the\countht}%
	\postscript{
		[ /Rect [ -1.5 -\linkdp.0 0\linkwd.0 0\linkht.5 ] 
		/Border [ 0 0 0 ]
		/Action << /Subtype /URI /URI (#2) >>
		/Subtype /Link
		/ANN pdfmark}{\rgb{1 0 0}{#1}}}
%
%
\else
\def\rgboo#1{\pdfliteral{#1 rg #1 RG}}

\pdfcatalog{/PageMode /UseOutlines}		
\def\bookmark#1#2{
	\pdfdest num \marknumber xyz
	\pdfoutline goto num \marknumber count #1 {#2}
	\advance\marknumber by1}
\def\pdfklink#1#2{%
	\noindent\pdfstartlink user
		{/Subtype /Link
		/Border [ 0 0 0 ]
		/A << /S /URI /URI (#2) >>}{\rgb{1 0 0}{#1}}%
	\pdfendlink}
\fi

\def\rgbo#1#2{\rgboo{#1}#2\rgboo{0 0 0}}
\def\rgb#1#2{\mark{#1}\rgbo{#1}{#2}\mark{0 0 0}}
\def\pdflink#1{\pdfklink{#1}{#1}}
\def\xxxlink#1{\pdfklink{[arXiv:#1]}{http://arXiv.org/abs/#1}}

\catcode`@=11

\def\wlog#1{}	


\def\makeheadline{\vbox to\z@{\vskip-36.5\p@
	\line{\vbox to8.5\p@{}\the\headline%
	\ifnum\pageno=\z@\rgboo{0 0 0}\else\rgboo{\topmark}\fi%
	}\vss}\nointerlineskip}
\headline={
	\ifnum\pageno=\z@
		\hfil
	\else
		\ifnum\pageno<\z@
			\ifodd\pageno
				\tenrm\romannumeral-\pageno\hfil\lefthead\hfil
			\else
				\tenrm\hfil\righthead\hfil\romannumeral-\pageno
			\fi
		\else
			\ifodd\pageno
				\tenrm\hfil\righthead\hfil\number\pageno
			\else
				\tenrm\number\pageno\hfil\lefthead\hfil
			\fi
		\fi
	\fi}

\catcode`@=12

\def\righthead{\hfil} \def\lefthead{\hfil}
\nopagenumbers


\def\chrulefill{\rgb{1 0 0}{\hrulefill}}
\def\cdotfill{\rgb{1 0 0}{\dotfill}}
\newcount\area	\area=1
\newcount\cross	\cross=1
\def\volume#1\par{\newpage\noindent{\biggest{\rgb{1 .5 0}{#1}}}
	\par\nobreak\bigskip\medskip\area=0}
\def\chapskip{\par\ifnum\area=0\bigskip\medskip\goodbreak
	\else\newpage\fi}
\def\chapy#1{\area=1\cross=0
	\xdef\lefthead{\rgbo{1 0 .5}{#1}}\vbox{\biggerer\offinterlineskip
	\line{\chrulefill¼\hphantom{\lefthead}\chrulefill}
	\line{\chrulefill¼\lefthead\chrulefill}}\par\nobreak\medskip}
\def\chap#1\par{\chapskip\bookmark3{#1}\chapy{#1}}
\def\sectskip{\par\ifnum\cross=0\bigskip\medskip\goodbreak
	\else\newpage\fi}
\def\secty#1{\cross=1
	\xdef\righthead{\rgbo{1 0 1}{#1}}\vbox{\bigger\offinterlineskip
	\line{\cdotfill¼\hphantom{\righthead}\cdotfill}
	\line{\cdotfill¼\righthead\cdotfill}}\par\nobreak\medskip}
\def\sect#1 #2\par{\sectskip\bookmark{#1}{#2}\secty{#2}}
\def\subsectskip{\par\ifdim\lastskip<\medskipamount
	\bigskip\medskip\goodbreak\else\nobreak\fi}
\def\subsecty#1{\noindent{\sectfont{\rgbo{.5 0 1}{#1}}}\par\nobreak\medskip}
\def\subsect#1\par{\subsectskip\bookmark0{#1}\subsecty{#1}}
\long\def\x#1 #2\par{\hangindent2\parindent%
\mark{0 0 1}\rgboo{0 0 1}{\bf Exercise #1}\\#2%
\par\rgboo{0 0 0}\mark{0 0 0}}
\def\refs{\bigskip\noindent{\bf \rgbo{0 .5 1}{REFERENCES}}\par\nobreak\medskip
	\frenchspacing \parskip=0pt \refrm \baselineskip=1.23em plus 1pt
	\def\ital##1Õ{{\refit##1\/}}}
\long\def\twocolumn#1#2{\hbox to\hsize{\vtop{\hsize=2.9in#1}
	\hfil\vtop{\hsize=2.9in #2}}}


\twelvepoint
\font\bigger=cmbx12 \sca2
\font\biggerer=cmb10 \sca5
\font\biggest=cmssdc10 scaled 4500
 \sca5

 \sca3


\def Ü{\relax\ifmmode\Rightarrow\else\expandafter\subsect\fi}
\def Û{\relax\ifmmode\Leftrightarrow\else\expandafter\sect\fi}
\def Ú{\relax\ifmmode\Leftarrow\else\expandafter\chap\fi}

\def\itemize#1 {\item{\bf#1}}
\def\itemizze#1 {\itemitem{\bf#1}}
\def\itemutem{\par\indent\indent \hangindent3\parindent \textindent}
\def\itemizzze#1 {\itemutem{\bf#1}}
\def ª{\relax\ifmmode\leftrightarrow\else\itemizze\fi}
\def Á{\relax\ifmmode\gets\else\itemizzze\fi}

\def\¢{\ominus}
\def\A{{\cal A}}  \def\B{{\cal B}}  \def\C{{\cal C}}  
         
      \def\M{{\cal M}}   \def\N{{\cal N}}  
\def\O{{\cal O}}  \def\P{{\cal P}}  \def\Q{{\cal Q}}

\def\Ä{\varphi}  \def\¿{\varpi}	\def\Ï{\vartheta}

\def ò{\relax\ifmmode\cdots\else\dotfill\fi}


\def\cvrule{\rgbo{0 .5 1}{\vrule}}
\def\chrule{\rgbo{0 .5 1}{\hrule}}
\def\boxit#1{\leavevmode\thinspace\hbox{\cvrule\vtop{\vbox{\chrule%
	\vskip3pt\kern1pt\hbox{\vphantom{\bf/}\thinspace\thinspace%
	{\bf#1}\thinspace\thinspace}}\kern1pt\vskip3pt\chrule}\cvrule}%
	\thinspace}
\def\Boxit#1{\noindent\vbox{\chrule\hbox{\cvrule\kern3pt\vbox{
	\advance\hsize-7pt\vskip-\parskip\kern3pt\bf#1
	\hbox{\vrule height0pt depth\dp\strutbox width0pt}
	\kern3pt}\kern3pt\cvrule}\chrule}}




\def\today{\ifcase\month\or
 January\or February\or March\or April\or May\or June\or July\or
 August\or September\or October\or November\or December\fi
 \space\number\day, \number\year}

\parindent=20pt
\newskip\normalparskip	\normalparskip=.7\medskipamount
\parskip=\normalparskip	



\catcode`\|=\active \catcode`\<=\active \catcode`\>=\active 
\def|{\relax\ifmmode\delimiter"026A30C \else$\mathchar"026A$\fi}
\def<{\relax\ifmmode\mathchar"313C \else$\mathchar"313C$\fi}
\def>{\relax\ifmmode\mathchar"313E \else$\mathchar"313E$\fi}


%
%
%
%
%
%
%

\def\thetitle#1#2#3#4#5{
 \def\titlefont{\biggest} \font\footrm=cmr10 \font\footit=cmti10
  \twelverm
	{\hbox to\hsize{#4 \hfill YITP-SB-#3}}\par
	\vskip.8in minus.1in {\center\baselineskip=2.2\normalbaselineskip
 {\titlefont #1}\par}{\center\baselineskip=\normalbaselineskip
 \vskip.5in minus.2in #2
	\vskip1.4in minus1.2in {\twelvebf ABSTRACT}\par}
 \vskip.1in\par
 \narrower\par#5\par\unnarrower\vskip3.5in minus3.3in\eject}
\def\paper\par#1\par#2\par#3\par#4\par#5\par{
	\thetitle{#1}{#2}{#3}{#4}{#5}} 
\def\author#1#2{#1 \vskip.1in {\twelveit #2}\vskip.1in}
\def\YITP{C. N. Yang Institute for Theoretical Physics\\
	State University of New York, Stony Brook, NY 11794-3840}
\def\WS{W. Siegel\footnote{$*$}{
	\pdflink{mailto:siegel@insti.physics.sunysb.edu}\\
	\pdfklink{http://insti.physics.sunysb.edu/\~{}siegel/plan.html}
	{http://insti.physics.sunysb.edu/\noexpand~siegel/plan.html}}}


\pageno=0

\paper

{\rgb{1 0.3 0}{AdS/CFT in superspace}}

\author\WS\YITP

10-16

May 13, 2010

We relate the 10D IIB superstring (primarily free supergravity) expanded about AdS$_5ð$S$^5$ to 4D N=4 Yang-Mills directly in superspace.  The sphere becomes the CFT internal space of projective superspace, plus a coordinate counting the number of supergluons.  The 32 fermionic coordinates of AdS become the 8 of the CFT in a spacecone gauge that manifestly preserves SO(3,1)$°$SO(4).  The PSU(2,2|4) algebra of AdS becomes the projective representation (plus the extra coordinate) on the boundary.  (A review of projective superspace and the projective lightcone limit is included for self-containment.)

\pageno=2

Û0 1. Introduction

Almost all of the work on the Anti de Sitter/Conformal Field Theory correspondence has involved an explicit Kaluza-Klein expansion over S$^5$, including the application of superspace methods [1].  
In fact, the use of 10D superfields (or even just 10D bosonic fields) can give simpler expressions than their Kaluza-Klein components for results such as bulk-to-bulk propagators [2].  It might even allow a more direct correspondence, relating first-quantization in the two theories.  Here we further elaborate on the relation between 10D IIB superspace and 4D N=4 projective superspace by showing how the coordinates of the 5-sphere appear in the 4D superspace.  The 10 coordinates decompose into the 4 spacetime + 4 internal coordinates of 4D N=4 projective superspace, the 1 ``holographic" coordinate whose endpoint defines the boundary, plus 1 extra coordinate that counts the Kaluza-Klein excitation level.

Concepts related to projective superspace have appeared in the literature:  (1)¼Projective spaces are a special case of ``flag spaces" that form rectangular matrices.   Here we focus on the case of square matrices, which is important both for invertibility and (pseudo)hermiticity.  (2) Projective spaces can also appear as the (usually) complex half of unitary Grassmannian cosets.  However, the full Grassmannian is often irrelevant:  For example, the Grassmannian U(4)/U(2)$^2$, after an appropriate Wick rotation, is relevant to the 4D conformal group, but has two real halves, parametrized by coordinates for both translations and conformal boosts, whereas only the former are desirable.  (Thus, we have sometimes referred to projective spaces as ``half-cosets".)  

(3) The harmonic superspace approach [3] can be treated as identical to the projective one Óon-shellÕ, where both use the same square of superspace coordinates.  However, projective superspace still uses the same coordinates off shell, whereas (except in the chiral case [4]) harmonic superspace requires additional internal coordinates that are invariant under the superconformal group [5].  The distinction is that the harmonic approach uses stronger boundary conditions in the internal space that are equivalent to the projective field equations.  In the AdS/CFT correspondence the distinction is usually moot, since first-quantized string theory puts external states on shell.  However, off-shell generalization will be useful for field theory, on both the AdS and CFT sides.  This has already proven to be the case for 4D N=2 supersymmetric theories [6] (for which projective superspace was first invented, before harmonic), where the additional coordinates of harmonic superspace were found redundant [7] (although they have proven useful for conceptual purposes, and deriving some results earlier).

Most of this paper will be a ``review" of previous papers [8], but presented in a more coherent way, with more details, and new results.  The following section is mostly a collection of definitions, conventions, and useful equations.  Section 3 treats off-shell 4D superfields with general spin by induced representations, which automatically introduce the extra bosonic coordinate associated with the 10D superstring on the boundary, but in a nondynamical way.  Section 4 considers the 4D field equations and their solution in terms of supertwistors, which allows a simple definition of the cut propagator, but requires careful handling of $i·$ prescriptions to get the St¬uckelberg-Feynman propagator.  

The actual AdS/CFT correspondence is discussed in section 5:  the supersymmetric definitions of the boundary limit and holography, their direct relation to the spacecone gauge, the 10D supergravity free field equations and their solution, and the role of the extra bosonic coordinate (balancing 9 for the 10D boundary vs.¼8 for 4D N=4 projective superspace).  As an example of the utility of the spacecone gauge, the scalar bosonic propagator on AdS$_5ð$S$^5$ is rederived by directly integrating the Klein-Gordon equation as a first-order differential equation in one variable (rather than the previous method of Weyl rescaling of the flat-space result).

Û3 2. Superspaces

Ü2.1. Full

Cosets are easier with classical groups.  The Poincar«e group is a contraction of a classical group, so the conformal group is easier; it's also more useful, since nonconformal theories can be treated as broken conformal ones.  In D=4 the superconformal group is (P)SU(N|2,2); to postpone considerations of reality properties, we'll Wick rotate to (P)SL(N|4) (corresponding to 2 space and 2 time dimensions).  Furthermore we can treat not only ``P" but also ``S" as gauge invariances rather than constraints; then an element of the group (or algebra) GL(N|4) is just an arbitrary real matrix (with appropriate grading).  Thus, no consideration of exponentiation or constraints on the coordinates is necessary.  The symmetry generators and covariant derivatives are then very simple:
$$ G_\M{}^\N = g_\M{}^\A »_\A{}^\N,ââD_\A{}^\B = (»_\A{}^\M)g_\M{}^\B $$
where $»_\A{}^\M=»/»g_\M{}^\A$, and we ordered the derivatives to the left in $D$ to keep grading signs trivial.  (The derivatives are meant to act only to the right of the $g$.  It's a kind of ``normal ordering".)  This corresponds to the usual identification of the symmetry generators $G$ acting to the left of the group coordinates $g_\M{}^\A$ and the covariant derivatives $D$ to the right, or the reverse for the inverse $g_\A{}^\M$.

The choice of gauge (isotropy) group is simply the choice of which constraints can be expressed linearly in covariant derivatives, instead of quadratically.  In principle all constraints can be expressed quadratically, but this tends to be awkward in general.  A simple example is the ordinary conformal group, with equations of motion quadratic in symmetry generators, which can be translated into the same for covariant derivatives (momentum, spin, and conformal weight) by factors of $g$ and $g^{-1}$.  They simplify because the covariant derivative for conformal boosts is set to vanish.  Also, the Lorentz and scale coordinates are generally replaced with spin and scale weight, giving them fixed ``values".

We now look at this construction in more detail, and generalize to supersymmetry.  Because of the use of GL(N|4) for D=4, this construction is simpler than using (the defining representation of) SO(D,2) for arbitrary D (for N=0), or the superconformal groups OSp(N|4) for D=3 or OSp*(8|2N) for D=6, since the latter require a quadratic constraint on the matrices.  The fact that the relevant cosets are projective spaces is a significant further simplification.

For a preliminary analysis, we divide up the graded matrices into their bosonic and fermionic parts:
$$ g_\M{}^\A = \bordermatrix{ & Ða & Ќ \cr
	Ñm & g_{Ñm}{}^{Ða} & g_{Ñm}{}^{Ќ} \cr
	е & g_{е}{}^{Ða} & g_{е}{}^{Ќ} \cr} $$

\noindent where barred Latin indices are bosonic internal GL(N) indices and barred Greek are fermionic spacetime GL(4) spinor indices, and then further divide the latter into 2 Lorentz GL(2) Weyl spinor indices, but reordered as determined by dimensional analysis (as is apparent when the individual coordinates/generators are identified): 
$$ g_\M{}^\A = \bordermatrix{ & Œ & Ða & ÀŒ \cr
	µ & g_µ{}^Œ & g_µ{}^{Ða} & g_µ{}^{ÀŒ} \cr
	Ñm & g_{Ñm}{}^Œ & g_{Ñm}{}^{Ða} & g_{Ñm}{}^{ÀŒ} \cr
	Àµ & g_{Àµ}{}^Œ & g_{Àµ}{}^{Ða} & g_{Àµ}{}^{ÀŒ} \cr} $$
$$ = \pmatrix { Lorentz + scale & supersymmetry & translation \cr
	S\hbox{-}supersymmetry & internal & supersymmetry \cr
	\hbox{Óconformal boostÕ} & S\hbox{-}supersymmetry & Lorentz - scale \cr} $$

\noindent Thus the scale weights (engineering dimensions) increase from lower-left to upper-right.
The usual full superspace is then obtained by gauging away the diagonal blocks, as well as the lower-left triangle (``lowering operators"), leaving only the coordinates for translations and supersymmetry:
$$ g_\M{}^\A £ \pmatrix{ 
	I & Ï_µ{}^{Ða} & x_µ{}^{ÀŒ} \cr
	0 & I & ÐÏ_{Ñm}{}^{ÀŒ} \cr
	0 & 0 & I \cr} $$

More choices can be obtained by also subdividing the N-valued internal indices, perhaps not equally, into n and N$-$n:
$$ g_\M{}^\A = \bordermatrix{ & Œ & a & a' & ÀŒ \cr
	µ & g_µ{}^Œ & g_µ{}^{a} & g_µ{}^{a'} & g_µ{}^{ÀŒ} \cr
	m & g_{m}{}^Œ & g_{m}{}^{a} & g_{m}{}^{a'} & g_{m}{}^{ÀŒ} \cr
	m' & g_{m'}{}^Œ & g_{m'}{}^{a} & g_{m'}{}^{a'} & g_{m'}{}^{ÀŒ} \cr
	Àµ & g_{Àµ}{}^Œ & g_{Àµ}{}^{a} & g_{Àµ}{}^{a'} & g_{Àµ}{}^{ÀŒ} \cr} $$

\noindent Again gauging away diagonal blocks and the lower-left triangle, we are left with an additional n(N$-$n) internal coordinates:
$$ g_\M{}^\A £ \pmatrix{ 
	I & Ï_µ{}^{a} & Ï_µ{}^{a'} & x_µ{}^{ÀŒ} \cr
	0 & I & y_{m}{}^{a'} & ÐÏ_{m}{}^{ÀŒ} \cr
	0 & 0 & I & ÐÏ_{m'}{}^{ÀŒ} \cr
	0 & 0 & 0 & I \cr} $$
For N=1 (``simple" superspace) this is identical to the previous case, but for N>1 it allows for generalizations that have proven necessary for most practical applications.  However, so far only N=2 superspace (``hyperspace") has been developed to a point approaching the usefulness of N=1.

Ü2.2. Projective as cosets

Projective spaces are obtained by gauging away parts of GL groups in the same manner as above (diagonal blocks + lower triangle), but dividing up the indices into only 2 parts.  So we reassemble the previous 4 parts, but differently than the 2 original blocks (bosonic + fermionic) as indicated by our reordering.  We then do a second reordering, as the standard bosonic + fermionic within each block:
$$ g_\M{}^\A = \bordermatrix{ & A & A' \cr
	M & g_M{}^A & g_M{}^{A'} \cr
	M' & g_{M'}{}^A & g_{M'}{}^{A'} \cr}
£ \pmatrix{ I & w_M{}^{A'} \cr 0 & I \cr}
= \bordermatrix{ & a & Œ & a' & ÀŒ \cr
	m & I & 0 & y_{m}{}^{a'} & ÐÏ_{m}{}^{ÀŒ} \cr
	µ & 0 & I & Ï_µ{}^{a'} & x_µ{}^{ÀŒ} \cr
	m' & 0 & 0 & I & 0 \cr
	˵ & 0 & 0 & 0 & I \cr} $$
This case has the same bosonic coordinates but half the anticommuting coordinates of the previous.  This is the smallest number of fermions we can get, since the gauge algebra must close, and we can't kill both a supersymmetry and its complex conjugate without killing translations.  This is useful for constructing actions, since $ÇdÏ=»/»Ï$ has positive mass dimension, so more $Ï$'s would require a Lagrangian lower in dimension.  For N=1, this is either chiral superspace (n=0, no $ÐÏ$'s), in the chiral representation, or antichiral (n=1, no $Ï$'s), in the antichiral representation.  For general N, n=0 is again chiral, and n=N antichiral, both with no $y$'s.  We also have a simple expression for the inverse matrix, in this gauge:
$$ g_\A{}^\M = \bordermatrix{ & M & M' \cr
	A & g_A{}^M & g_A{}^{M'} \cr
	A' & g_{A'}{}^M & g_{A'}{}^{M'} \cr}
£ \pmatrix{ I & -w_A{}^{M'} \cr 0 & I \cr} $$
where this matrix $w$ is the same as the previous.  (Symmetry and gauge indices lose their distinction after gauge fixing.)

As we'll discuss in detail later, only the case n=N/2 (and thus even N) allows real superfields, since only it makes $w$ a square matrix, with equal range for the $A$ index and its ``charge conjugate" $A'$.  This is especially clear if we note that it's the only case where there are equal numbers of $Ï$'s and $ÐÏ$'s.  (Of course, the full superspaces also allow real superfields.)  Since this makes them the most useful, we'll often use the term ``projective" to refer to them specifically.

We now derive the form of the symmetry generators and covariant derivatives before gauge fixing, in a convenient coordinate representation, using matrix methods.  We write in matrix notation
\vskip-.1in
$$ g = 
\pmatrix{ I & w \cr 0 & I \cr}\pmatrix{ u & 0 \cr 0 & Ðu{}^{-1} \cr}\pmatrix{I & 0 \cr -v & I\cr}
= \pmatrix{ u - wÐu{}^{-1}v & wÐu{}^{-1} \cr -Ðu{}^{-1}v & Ðu{}^{-1} \cr} $$

\noindent which defines the coordinates $w_M{}^{M'}$, $u_M{}^A$, $Ðu_{A'}{}^{M'}$, and $v_A{}^{A'}$.  Note that in this representation we have (using $sdet(XY)=sdet(X)sdet(Y)$)
$$ sdet¼g = {sdet¼u\over sdet¼Ðu} $$

It's actually easier to derive the generators from the form of finite transformations, rather than using $G=g»_g$ and $D=(»_g)g$ and then using the above redefinitions of the elements of $g$ in terms of $w,u,Ðu,v$.
Writing the symmetry transformation in terms of
$$ g_0 = \pmatrix{ a & b \cr c & d \cr},âg_0^{-1} = \pmatrix{ ÷d & -÷b \cr -÷c & ÷a \cr} $$

\noindent in $g'=g_0 g$ and $g'^{-1}=g^{-1}g_0^{-1}$ (whichever is simpler), we find the finite superconformal transformations
$$ u' = (w÷c+÷d)^{-1}u,ââÐu' = Ðu(cw+d)^{-1}, $$
$$ w' = (aw+b)(cw+d)^{-1} = (w÷c+÷d)^{-1}(w÷a+÷b), $$
$$ v' = v -Ðu(cw+d)^{-1}cu = v -Ðu÷c(w÷c+÷d)^{-1}u $$

Continuing to use matrix notation, we can write the infinitesimal transformations as
$$ ¶ = str[ (¶w)»_w + (¶u)»_u + (¶Ðu)»_{Ðu} + (¶v)»_v ] $$
using $w,u,Ðu,v$ as indices labeling the blocks of the derivatives.  This allows cycling all parameters inside the supertrace to the far left, again using the ``normal-ordering" convention that derivatives are understood to act only to the right of everything in the generators.  (We can do the same for $D$ by right multiplication.)  We thus have
\vskip-.2in
$$ G = g»_g = \pmatrix{ w»_w +u»_u & -w»_w w -u»_u w -w»_{Ðu}Ðu -u»_v Ðu \cr
	»_w & -»_w w -»_{Ðu}Ðu \cr} 
	­ \pmatrix{ G_u & -G_v \cr G_w & -G_{Ðu} \cr} $$
$$ D = »_g g = \pmatrix{ »_u u +»_v v & -»_v \cr
	Ðu »_w u +v»_u u +Ðu »_{Ðu} v +v»_v v & -Ðu »_{Ðu} -v»_v \cr} 
	­ \pmatrix{ D_u & -D_v \cr D_w & -D_{Ðu} \cr} $$

\noindent (introducing some convenient signs by convention).

Ü2.3. Projective by projection

The interesting properties of these cases follow from the fact that the coset coordinates fit into a rectangle.  Furthermore, although the full, ``left" index is required for manifest symmetry, the gauge group necessarily breaks the ``right" index into 2 pieces.  We can therefore begin with a rectangle that keeps the full left index, but only the part of the right index that contains the coset:
$$ g_\M{}^\A £ Ðz_\M{}^{A'} = \bordermatrix{ & A' \cr
	M & z_M{}^{A'} \cr
	M' & z_{M'}{}^{A'} \cr} $$
And we can do the analogous for the inverse group element:
$$ g_\A{}^\M £ z_A{}^M = \bordermatrix{ & M & M' \cr
	A & z_A{}^M & z_A{}^{M'} \cr} $$
Then all that's left of the relation between the group element and its inverse is the orthogonality relation
$$ z_A{}^\M Ðz_\M{}^{A'} = 0 $$
Furthermore, all that's left of the original gauge invariance is the block diagonal pieces, one of which acts only on $z$ (GL(n|2) for the superconformal group), and the other only on $Ðz$ (GL(N$-$n|2)).  Note that neither $z$ nor $Ðz$ contains the coordinates for conformal boosts.

As for other projective spaces (cf.¼RP(n) and CP(n); the chiral case here is HP($ü$N|1)), the surviving coordinates $w$ can be defined in a gauge-invariant way, which is a simpler way to see their symmetry transformations.  An easy way to do this is by solving the orthogonality condition, as
$$ Ðz_\M{}^{A'} = ( w_M{}^{N'} , ¶_{M'}^{N'} ) Ðu_{N'}{}^{A'} ,ââ
	z_A{}^\M = u_A{}^N ( ¶_N^M , -w_N{}^{M'} ) $$
which reproduces some of the coordinates of the coset formulation.  Specifically, we can identify these if we write in matrix notation
$$ Ðz = \pmatrix{ w \cr I \cr}Ðu{}^{-1},ââz = u^{-1}\pmatrix{ I & -w \cr} $$
Only $u$ and $Ðu$ transform under their respective gauge transformations.  This defines $w$ as the ``ratio" of the 2 blocks of either $z$ or $Ðz$:
$$ w_M{}^{M'} = Ðz_M{}^{A'}(Ðz_{M'}{}^{A'})^{-1} = -(z_A{}^M)^{-1}z_A{}^{M'} $$
where the inverses are matrix inverses of those blocks.  

The symmetry transformation of $w$ then follows as a ``fractional linear" (``projective") transformation:  As for the coset case,
$$ Ðz' = g_0 Ðz,âz' = zg_0^{-1}âÜâw' = (aw+b)(cw+d)^{-1} = (w÷c+÷d)^{-1}(w÷a+÷b) $$
A special case is ordinary conformal symmetry (N=0), where all the above are 2$ð$2 matrices:  This takes a simpler form than in the usual vector notation, just as for the case of SO(3,1) on 2D Euclidean space.  Here the simplification arises from using quaternions instead of 4D vectors, while in the 2D case it was complex numbers in place of 2-vectors.  

From the same derivation we also have the transformations of the $u$'s:  Again as from the coset treatment,
$$ Ðu' = Ðu(cw+d)^{-1},ââu' = (w÷c+÷d)^{-1}u $$
Note that in the gauge (or subject to the constraint) $sdet¼g=1$, we have $sdet¼u=sdet¼Ðu$; thus
$$ sdet(g_0) = 1âÜâsdet(cw+d) = sdet(w÷c+÷d) $$

We can also construct symmety invariants in a similar way to (and implied by) the coset construction (consider $g^{-1}dg$ and $g_2^{-1}g_1$), as differentials or finite differences:
$$ z_A{}^\M dÐz_\M{}^{A'} = u_A{}^M (dw_M{}^{M'}) Ðu_{M'}{}^{A'},ââ
	z_{2A}{}^\M Ðz_{1\M}{}^{A'} = u_{2A}{}^M (w_1-w_2)_M{}^{M'} Ðu_{1M'}{}^{A'} $$
The $u$'s are pure gauge; symmetry- and gauge-invariant quantities depend only on differentials or differences of $w$, according to the translation (``$b$") part of the symmetry.  These translations include the usual spacetime ones, some of the internal symmetry, and half the supersymmetries (as in the special case of chiral superspace).

The form of the symmetry generators in terms of $w$ and $u$ can again easily be derived from the finite forms of the transformations (taking the infinitesimal limit).  We thus find the basis
$$ G_w = »_w,ââG_u = w»_w + u »_u,ââG_{Ðu} = »_w w + »_{Ðu}Ðu ,ââ
	G_v = w »_w w + u »_u w + w »_{Ðu}Ðu $$
which are the coset-space generators less the $»_v$ term in $G_v$.  One can also check that these operators are permuted by the ``inversion" (a particular case of the above finite transformations)
$$ g_0 = \pmatrix{ 0 & -I \cr I & 0 \cr} :ââw £ -w^{-1},ââu £ w^{-1}u,ââ Ðu £ Ðuw^{-1} $$

Although the covariant derivatives $D_u$ and $D_{Ðu}$ for the gauge group are obvious from the way they act on the group indices,
$$ D_u = »_u u,ââD_{Ðu} = Ðu»_{Ðu} $$
the remaining derivatives $D_w$ can't be found that commute with the symmetry generators $G_v$.  However, we can define
$$ D_w = Ðu »_w u $$
that commute with all but $G_v$.  This is the usual procedure for ordinary conformal symmetry, where coordinates are not introduced for conformal boosts, so (``covariant") translational derivatives don't commute with them.

\newpage
Û3 3. Off-shell superfields

Ü3.1. Induced representations

To generalize our discussion of superspaces to include spin, we begin by reviewing the general construction of Hilbert spaces for cosets.  We first define a vacuum state invariant under the gauge group:  In terms of its generators $H_î$,
$$ H_î |0Ô = Ò0|H_î = 0 $$
(For some purposes, we can think of the gauge generators as ``lowering operators".  In general, we don't need a Hilbert space for this construction, but only a vector space; the bras then form the dual space to the kets.)  A coordinate basis for the coset can then be defined, for arbitrary group coordinates $Œ$, as
$$ |ŒÔ = g(Œ)|0Ô,ââҌ| = Ò0|g^{-1}(Œ) $$
(where $g(0)=I$) and thus invariant under a gauge transformation
$$ g'|0Ô ­ gh|0Ô = g|0Ô $$
for arbitrary element $h(Œ)$ of the gauge group.  The wave function is then defined with respect to this basis as
$$ Æ(Œ) ­ Ҍ|ÆÔ = Ò0|g^{-1}(Œ)|ÆÔ $$
from which it follows that its covariant derivative with respect to the gauge group (but not the coset) vanishes:
$$ -D_î Æ(Œ) = Ò0|H_î g^{-1}(Œ)|ÆÔ = 0 $$
On the other (right) hand, the symmetry generators act in the expected way:
$$ -G_I Æ(Œ) = Ò0|g^{-1}(Œ)öG_I|ÆÔ = (öG_I Æ)(Œ) $$
(using a $ß{\phantom G}$ on $G$ to distinguish the Hilbert-space operator from the differential operator:  $öG_I=(T_i,H_î)$.).

This is sufficient for coordinate representations.  But usually in quantum mechanics we want to consider more general representations by adding ``spin" to such ``orbital" generators.  This is accomplished by first introducing spin degrees of freedom, and then tying them to the group by modifying the gauge-group constraints.  
So we first introduce a basis $|{}^AÔ$ (and its dual $Ò{}_A|$) for a matrix representation $÷H_î$ for the gauge group, 
$$ Ò{}_A|H_î = ÷H_{îA}{}^BÒ{}_B|,ââH_î|{}^AÔ = |{}^BÔ÷H_{îB}{}^A $$
(We use different index notation here from the rest of the paper due to shortage of alphabets.)  We then define a basis for the Hilbert space by using this gauge group basis as our new (degenerate) vacuum,
$$ |{}^A,ŒÔ ­ g(Œ)|{}^AÔ $$
to get the generalizations of the previous
$$ Æ_A(Œ) ­ Ò{}_A,Œ|ÆÔâÜâ-D_î Æ_A(Œ) = ÷H_{îA}{}^B Æ_B(Œ),ââ-G_I Æ_A(Œ) = (öG_I Æ)_A(Œ) $$
The wave function now depends also on the gauge-group coordinates, but this dependence is fixed independent of the state:  For example, in the 2-exponential coordinate system $Œ^I=(º^i,©^î)$ where dependence on the coset generators $T_i$ is explicitly factorized,
$$ Æ_A(Œ) = Ò{}_A|e^{-i©^î H_î}e^{-iº^i T_i}|ÆÔ 
	= (e^{-i©^î ÷H_î})_A{}^M Ò{}_M|e^{-iº^i T_i}|ÆÔ ­ e_A{}^M(©)Æ_M(º) $$
where $e_A{}^M$ is a ``vielbein" depending on only the gauge coordinates $©$, and can be gauged to the identity, while $Æ_M$ depends on only the coset coordinates $º$. 
Since we know $D$ in terms of derivatives, $D_î=-÷H_î$ can be solved to replace partial derivatives with respect to gauge-group coordinates with matrices, in both $D_I$ and $G_I$.

The commutation relations of the surviving covariant derivatives
$$ [ D_i, D_j Õ = f_{ij}{}^k D_k + f_{ij}{}^û D_û $$
then identify $f_{ij}{}^k$ as the ``torsion", while $f_{ij}{}^û$ is the ``curvature".

This approach is effectively what is done in the usual analysis of the ordinary conformal group, or just the Poincar«e group:  For example, in Wigner's analysis of 4D Poincar«e representations, the spin is defined essentially as the covariant derivative left over when orbital angular momentum is subtracted from the full Lorentz generators.  The Pauli-Luba«nski equation (as well as the Klein-Gordon), expressed in terms of group generators, then directly reduces to covariant derivatives.  

Ü3.2. Spin

In our approach we keep $D_v=0$ unmodified, since it's automatic in the projective description, but introduce spin to replace $D_u$ and $D_{Ðu}$.  (Of course, $D_w$ is not in the gauge group.)  Then
$$ D_u ­ »_u u = s_u ­ u^{-1}ös_u u,ââD_{Ðu} ­ Ðu »_{Ðu} = s_{Ðu} ­ Ðu ös_{Ðu} Ðu{}^{-1}âÜ $$
$$ G_w = »_w,ââG_u = w»_w +ös_u,ââG_{Ðu} = »_w w +ös_{Ðu},ââ
	G_v = w»_w w +ös_u w +wös_{Ðu} $$
where the $ös$'s are defined to act on ``curved" indices $M,M'$ rather than ``flat" indices $A,A'$.  

Our flat/curved terminology is by analogy to general relativity, where ``flat" indices carry the Lorentz gauge symmetry, and are how spin is introduced, while ``curved" indices, and the coordinates that carry them, are acted on by any global symmetry of the space under consideration.  In fact, in the bosonic case our gauge group GL(2)$°$GL(2) is just the Lorentz group, scale transformations (for which the ``spin" part is the scale weight), and the purely gauge GL(1) that reduces GL(4) to the (Wick-rotated) conformal group SL(4).

Since our gauge group is GL(n|2)$°$GL(N-n|2), it's clear how this works:  The gauge generators $D_u$ and $D_{Ðu}$ carry flat indices; their irreducible matrix representations carry arbitrary mixtures of these defining indices, up and down, with arbitrary graded (anti)symmetrizations (but with arbitrary values of the Abelian GL(1) charges, and maybe some supertrace conditions).  Thus our original fields $\on\circ ì$ carry these flat indices, are scalars with respect to the symmetry group, and satisfy the constraints $D_u-s_u=D_{Ðu}-s_{Ðu}=0$.  But we can explicitly solve these constraints in terms of fields $ì$ that carry only curved indices, by using $u$ and $Ðu$ as ``vielbeins" to convert flat indices into curved.  The fields with curved indices then depend only on $w$, and are gauge invariant, but are no longer scalars: The $ös$'s in $G$ act the same way on the curved indices as the $s$'s acted on the flat (and themselves carry curved indices).

It's sufficient to consider an example with one of each type of index, primed and unprimed (up vs.¼down indices should be obvious):
$$ s_A{}^C \on\circ ì_{B'}{}^D = ¶_A^D \on\circ ì_{B'}{}^C -r ¶_A^C \on\circ ì_{B'}{}^D,ââ
	s_{A'}{}^{C'}\on\circ ì_{B'}{}^D = 
		¶_{B'}^{C'}\on\circ ì_{A'}{}^D - Ðr ¶_{A'}{}^{C'} \on\circ ì_{B'}{}^D $$
(with extra signs from index reordering implicit) where $r+Ðr$ is the superscale weight (see below) and $str¼s-str¼Ðs$ (the ``$-$" comes from the definition of $D_{Ðu}$ and $G_{Ðu}$) is related to the super-(internal-)U(1) charge (or superhelicity; see the following section).  The solution to the constraints is
$$ \on\circ ì_{A'}{}^A(w,u,Ðu) = (sdet¼u)^{-r-Ðr} Ðu_{A'}{}^{M'}ì_{M'}{}^{M}(w) u_M{}^A,ââ
	\on\circ ì{}'_{A'}{}^A(w,u,Ðu) = \on\circ ì_{A'}{}^A(w',u',Ðu') $$
$$ Üâì'_{M'}{}^{M}(w) = 
	[sdet(cw+d)]^{r+Ðr}(cw+d)^{-1}{}_{M'}{}^{N'}ì_{N'}{}^{N}(w')(w÷c+÷d)^{-1}{}_N{}^M $$
where $r$ and $Ðr$ appear only in the combination $r+Ðr$ because we have used the ``S" constraint on $g$, $sdetÊu=sdetÊÐu$ (which implies the analogous on $g_0$, $sdet(w÷c+÷d)=sdet(cw+d)$).

In the physically most interesting cases (e.g., the N=2 scalar multiplet or the N=4 vector multiplet) the field strength $\on\circ ì$ is a scalar (and $r+Ðr$ is nonzero).  It then depends on only $w$ and the ``extra coordinate" $sdetÊu$.  This is well known from the nonsupersymmetric case, where this extra coordinate is related to the $x_0$ of the projective lightcone.

A linear form of transformation on indices can be obtained by using $z$ and $Ðz$ to convert flat indices into full GL(N|4) curved indices; e.g.,
$$ ì_\M{}^\N ¾ Ðz_\M{}^{A'}ì_{A'}{}^A z_A{}^\N $$
But such fields are constrained,
$$ z_A{}^\M ì_\M{}^\N = ì_\M{}^\N Ðz_\N{}^{A'} = 0 $$
Solving the constraints leads back to the above fields and yields their nonlinear transformations.

Note that the fermionic part of the spin is usually assumed to vanish, in agreement with known physical examples.  This implies that their superpartners do also, so in those cases $s$ vanishes except for the chiral case, where only $s_u$ (consisting of just $s_Œ{}^º$) is nonvanishing, or the antichiral case, where only $s_{Ðu}$ is.

Another way to account for the superscale weight is to define a field $ì$ to be a density by requiring that $dw¼ì^{1/¿}$ transform as a scalar,
$$ dw¼ì'^{1/¿}(w) = dw'¼ì^{1/¿}(w') $$
where ``$dw$" is the naive integration measure over all the components of $w$, so $ì$ is a density of (superscale) weight ``$¿$".  (We may have switched active vs.¼passive transformations.)  The transformation law for $dw$ can be found from $dÐz$ (or $dz$), which is invariant because $sdet(g)=1$.  (The relation of the superdeterminant to Jacobians, as the generalization of the bosonic case, follows from its definition in terms of a Gaussian integral.)  This is true already for the part of the measure $dÐz$ coming from any one particular value of $A'$ in $Ðz_\M{}^{A'}$.  We then separate out $dw$ and $dÐu$ in $Ðz=(w,I)Ðu{}^{-1}$:
$$ dÐz_\M{}^{A'} = (dw_M{}^{N'},0)Ðu_{N'}{}^{A'} +(w_M{}^{N'},¶_{M'}^{N'})dÐu_{N'}{}^{A'} $$
$$ ÜâdÐz = dwÊ(sdet¼Ðu)^{-strÊI_u} ð d(Ðu^{-1}),ââstrÊI_u = n - 2 $$
where the exponent comes from multiplying the contributions from each particular value of $M$.  The superconformal transformation of $d(Ðu{}^{-1})$ then follows from that of $Ðu{}^{-1}$ by a similar manipulation:
$$ d(Ðu{}^{-1})' = d(Ðu{}^{-1})[sdet(cw+d)]^{strÊI_{Ðu}},ââstrÊI_{Ðu} = (N-n) -2 $$
$$ Üâdw' = dwÊ[sdet(cw+d)]^{-strÊI},ââstrÊI = N-4 $$
(This derivation is thus singular for N=4, related to the additional ``P" gauge invariance.)  The superconformal transformation of $ì$ is then
$$ ì'(w) = [sdet(cw+d)]^{-¿ÊstrÊI}ì(w'),ââw' = (aw+b)(cw+d)^{-1} $$
We can identify $-¿ÊstrÊI$ with $r+Ðr$ of the previous derivation.  ($r+Ðr$ needn't vanish for N=4, where $dw$ is a scalar.)

Ü3.3. Charge conjugation

As explained previously, only the cases N=2n, where $w$ is square, allow the existence of real superfields.  Because of the Wick rotation used to conveniently describe the superconformal group, fields will satisfy nontrivial reality conditions.  We really don't need to Wick rotate:  If you ignore reality, it doesn't make a difference; just treat any variable and its complex conjugate as algebraically independent.  (However, there can be some topologcial complications, which we'll ignore, at least for now.)  Reality for the superconformal group is expressed as a pseudo-unitarity condition (the ``U" in (P)SU(N|2,2)),
$$ gÿçg = gçgÿ = ç,ââç^2 = I,ââçÿ = ç;ââ
	ç^{À{\M}\N} = \bordermatrix{ & n & Ã & n' & ÀÃ \cr
	Àm & I & 0 & 0 & 0 \cr
	˵ & 0 & 0 & 0 & -iC \cr
	Àm' & 0 & 0 & I & 0 \cr
	µ & 0 & iC & 0 & 0 \cr} $$
in terms of the SL(2) and U(2) metrics, e.g.,
$$ C^{µÃ} = \pmatrix{ 0 & i \cr -i & 0 \cr},ââI^{Àmn} = ¶_m^n $$

It isn't useful to solve for the reality conditions on the components of $g$ because of the nonlinearity, and because some of the complex conjugates of components of $w$ are in $v$.  (So we have chosen a complex gauge by eliminating $v$.)  Instead, we use this unitarity condition to define ``charge conjugates" of elements of $g$ that transform in the same way under the symmetry group, although differently under the gauge group.  Specifically, we need this only for the coset:
$$ \C (w') = (\C w)' $$
where $\C$ acts on $w'$ as if it were $w$, and $'$ acts on $\C w$ as if it were $w$; thus superconformal transformations and charge conjugation commute.
We therefore need to use only the fact that the symmetry transformation $g_0$ used in $g'=g_0 g$ satisfies the same unitarity condition as $g$ above.  This fact can then be applied as well to the transformations on the projective space, $Ðz'=g_0 Ðz$ and $z'=zg_0^{-1}$.  The goal will be to define a charge conjugation $\C$ of fields that involves their complex (hermitian) conjugation, but still gives fields that depend on $w$ (and not $wÿ$, whatever that is).  Thus for flat superfields
$$ (\C \on\circ ì)(w,u,Ðu) ­ [\on\circ ì( \C w,\C u,\C Ðu)]ÿ $$
where ``$\C w$" is some function of $wÿ$ (so $ìÿ$ gives back $w$) that transforms the same as $w$ under superconformal transformations.  The relation for curved superfields then follows.  For real fields (when they can be defined), $\C ì$ is identified with $ì$.

We thus define the action of charge conjugation $\C$ on the coordinates by
$$ \C g ­ gçßç = ç(g^{-1})ÿßç,ââ
	ßç^{À{\A}\B} = \bordermatrix{ & B & B' \cr ÀA & 0 & -I \cr ÀA' & I & 0 \cr} $$
In the former form the symmetry transformation is obvious, while in the latter form $ç$ mixes only the symmetry indices, with $ßç$ chosen to mix the gauge indices to relate the pieces appearing in the projective approach:
$$ (\C Ðz_\M{}^{A'})ÿ = -z_A{}^\N ç_{\NÀ{\M}},ââ
	(\C z_A{}^{\M})ÿ = ç^{À{\M}\N} Ðz_\N{}^{A'} $$
relating $z$ to the complex conjugate of $Ðz$.  (The ``$-$" sign, from $ßç$, preserves $sdetÊg=1$.)  The gauge indices don't match because charge conjugation switches primed and unprimed indices; but $w$ is gauge invariant.  We could match indices by putting back the identities in $ßç$; for the example of the previous subsection, the flat field would then satisfy
$$ (\C\on\circ ì_{A'}{}^A)(w,u,Ðu) ­ ßç_{A'ÀB}[\on\circ ì( \C w,\C u,\C Ðu)]ÿ{}^{ÀB}{}_{ÀB'}öç^{ÀB'A} $$

Independent of coordinate choice, we find as a result
$$ (\C G)ÿ = -ç^{-1}Gç,ââ(\C D)ÿ = -ßçÿDßçÿ^{-1} $$
(but we have chosen $ç^{-1}=ç$, $ßç^{-1}=ßçÿ=-ßç$).  More explicitly, and taking into account (i.e., undoing) that the above hermitian conjugation includes matrix transposition,
$$ \C :ââD_w £ -D_w,âD_v £ -D_v,âD_u ª -D_{Ðu} $$

We then find the conjugation of $w$, which we can write as
$$ (\C w)ÿ^{ÀM'}{}_{ÀN} = \bordermatrix{ & Àn & ÀÃ \cr
	Àm' & -y^{-1}{}_{m'}{}^n & -iy^{-1}{}_{m'}{}^n ÐÏ_n{}^{Àµ}C_{ÀµÀÃ} \cr
	µ & -iC^{µÃ}Ï_Ã{}^{n'}y^{-1}{}_{n'}{}^n 
			& -C^{µÃ}(x_Ã{}^{Àµ} -Ï_Ã{}^{n'}y^{-1}{}_{n'}{}^n ÐÏ_n{}^{Àµ})C_{ÀµÀÃ} \cr} $$

\noindent (For N=0, $\C x$ is just $x$:  The factors of $C$ in its hermitian conjugation are because it's $x^{µÀµ}$ that's hermitian.  Note that $C_{µÃ}=-C^{µÃ}$.)  For deriving charge conjugation for spin, it's also useful to have
$$ (\C u)ÿ = Ðu Ð{\A}^{-1}(w),ââ(\C Ðu)ÿ = \A^{-1}(w) u $$
$$ \A_{MÀN'} = \bordermatrix{ & Àn' & Ã \cr
		m & y_m{}^{n'} & 0 \cr
		µ & Ï_µ{}^{n'} & -iC_{µÃ}\cr},ââ
	Ð{\A}^{ÀMN'} = \bordermatrix{ & n' & ÀÃ \cr
		Àm & y_m{}^{n'} & ÐÏ_m{}^{ÀÃ} \cr
		Àµ & 0 & -iC^{ÀµÀÃ} \cr} $$
		
$$ ÜâsdetÊ\A = sdetÊÐ{\A} = det¼y $$
For the same example, we then have
$$ (\C ì_{M'}{}^M)(w) = (det¼y)^{r+Ðr} Ð{\A}^{-1}_{M'ÀN}[ì (\C w)]ÿ{}^{ÀN}{}_{ÀP'}\A^{-1ÀP'M}  $$

Another way to generalize to nonvanishing superscale weight is by considering densities, as for superconformal transformations.
We then need to relate $d(\C w)ÿ$ to $dw$:  With the help of the identity
$$ sdet(e^X) = e^{strÊX}âÜâsdet(X°Y) = (sdetÊX)^{strÊI_Y}(sdetÊY)^{strÊI_X} $$
to handle the two indices on $-d(y^{-1})=y^{-1}(dy)y^{-1}$, we find
$$ [d(\C w)]ÿ = dwÊ(det¼y)^{-strÊI} $$
Thus, requiring $dw¼ì^{1/¿}$ act as a scalar under charge conjugation,
$$ dwÊ[(\C ì)(w)]^{1/¿} ­ Ód(\C w)[ì(\C w)]^{1/¿}Õÿ $$
$$ Üâ(\C ì)(w) = (det¼y)^{-¿ÊstrÊI}[ì(\C w)]ÿ $$

Invariance of an action under charge conjugation (e.g., the known cases for N=2) implies its reality, since it has no explicit dependence on the coordinates.

\newpage
Û3 4. On-shell superfields

Ü4.1. Field equations

We have already seen that the group-space representation can be conveniently reduced by the use of constraints linear in the covariant derivatives, which define a coset space, of lower dimension.  Another way to reduce a representation is by field equations.  These are normally quadratic in the covariant derivatives, and thus can't be solved algebraically.  (However, we'll find a lightcone/supertwistor solution below.)  These equations apply only to field strengths.

These constraints, before introduction of the gauge group, carry the full range of indices, and thus can be written in terms of either the covariant derivatives or the symmetry generators, by virtue of the relation
$$ G_\M{}^\N = g_\M{}^\A D_\A{}^\B g_\B{}^\N $$
We can then consider the possible reduction of $DD$ or the equivalent $GG$ constraints simply by graded (anti)symmetrization and supertracing of the 4 indices.  The possible choices vary according to the number of spacetime dimensions in the final result:  One choice gives the desired 4D Minkowski space; another gives 5D anti-de Sitter space (where (P)SU(N|2,2) is the super anti-de Sitter group); yet another (in the case N=4) gives the 10D space AdS${}_5ð$S${}^5$, relevant for the AdS/CFT correspondence.

Restricting ourselves to the 4D case, the result can be obtained by noting that it is the tensor equation [9]
$$ G_{(\M}{}^{(\P}G_{\N]}{}^{\Q]} = 0¼mod¼¶¼terms $$
that includes the massless Klein-Gordon equation $p^2=0$.  
The equation is determined only up to Kronecker $¶$ terms, which don't contribute to the Klein-Gordon equation, and has this ambiguity because of the gauge invariance
$$ G_{\M}{}^{\N} £ G_{\M}{}^{\N} + ¶_{\M}^{\N}A $$
for arbitrary operator $A$.  (Because it's Abelian, this is the same gauge symmetry as in the  gauge group generated by $D$'s; ``Abelian" means it can be considered as either left or right.)

The bosonic case (N=0) is also a special case of the bosonic conformal equations in D arbitrary dimensions [10]
$$ G_{(\un M}{}^{\un P}G_{\un N) \un P} - tr = 0 $$
in terms of (D+2)-valued vector indices $\un M$ for the SO(D,2) generators.  These include the Klein-Gordon equation, the general field equation for all spin, and constraints on the conformal weight and spin.

The supersymmetric equation of motion also includes the general (massless) supersymmetry free field equation $Öpq=0$, the Pauli-Luba«nski equation, several $qq$ equations often seen in supersymmetry (usually as $dd$), equations involving the internal symmetry generators, and various redundant equations.

The number of field equations we wrote above in terms of the symmetry generators is reduced by the gauge constraints.  In terms of these generators, many equations are redundant; alternatively, we can start with the equations written in terms of the covariant derivatives, where some drop out automatically.  Either way, the net result is that the equations on projective space reduce to (all mod $¶$ terms for the $s$'s)
$$ »_{(M'}{}^{(P} »_{N']}{}^{Q]} = s_M{}^{(P} »_{N'}{}^{Q]} = s_{(M'}{}^{P'} »_{N']}{}^Q = 0 $$
$$ s_M{}^P s_{N'}{}^{Q'} = s_{(M}{}^{(P} s_{N]}{}^{Q]} = s_{(M'}{}^{(P'} s_{N']}{}^{Q']} = 0 $$
The first set of equations is for arbitrary massless representations of supersymmetry, the second set restricts the index structure for specialization to conformal supersymmetry.  (A similar separation can be made for the bosonic case in arbitrary dimensions.)  Specifically, the second set places the restriction that superconformal representations have only primed or only unprimed indices, and fixes the value of the superscale weight.

The list of the spin-free part of these reduced equations is:
$$ »_x »_x = »_x »_Ï = »_Ï »_Ï = »_Ï »_{ÐÏ} + »_x »_y = »_y »_Ï = »_y »_y = 0 $$
(and complex conjugates).  Internal indices are symmetrized, while Weyl spinor indices are contracted (antisymmetrized).  The $»_y$-free equations should be familiar from N=1 chiral scalars:  They include the Klein-Gordon, Weyl spinor, and auxiliary field equations, respectively.  The equation with all types of derivatives (and thus 2 different types of terms, each with only 1 of each kind of index, and thus no symmetrization possible) shows that any $y$-dependent term shows up without $y$ at higher order in $Ï$ and $ÐÏ$ with $x$-derivatives, and that all terms with both $Ï$ and $ÐÏ$ are of this form.  

Taylor expansion is sufficient for the $y$'s, since setting both primed indices equal and both unprimed indices equal in the $»_y »_y$ equation says the field is linear in each $y$.  (Of course we can always Taylor expand in the $Ï$'s.)  Then the non-$»_x$ equations say that all component fields in this Taylor expansion in $y$'s and $Ï$'s are totally antisymmetric in unprimed internal indices and separately also in primed.

We now examine the component expansion for N=4, n=2.  The result is straightforward:
$$ ì = (Ä + y_m{}^{m'} Ä_{m'}{}^m +üy^2 ÐÄ) + Ï_µ{}^{m'}(Â_{m'}{}^µ +y_{m'}{}^m Â_m{}^µ)
	+ ÐÏ_m{}^{Àµ}(ÐÂ_{Àµ}{}^m +y_{m'}{}^m ÐÂ_{Àµ}{}^{m'}) $$
$$ + (Ï^2_{µÃ}f^{µÃ} +ÐÏ^{2ÀµÀÃ}Ðf_{ÀµÀÃ}) 
	-i Ï_µ{}^{m'}ÐÏ_m{}^{Àµ}»_{Àµ}{}^µ (Ä_{m'}{}^m +y_{m'}{}^m ÐÄ) $$
$$ -iÏ^2_{µÃ}ÐÏ_m{}^{Àµ}»_{Àµ}{}^µ Â^{mÃ} -iÐÏ^{2ÀµÀÃ}Ï_µ{}^{m'}»_{Àµ}{}^µ ÐÂ_{ÀÃm'} 
	-Ï^2_{µÃ}ÐÏ^{2ÀµÀÃ}»_{Àµ}{}^µ »_{ÀÃ}{}^à ÐÄ $$
where we have used the internal SL(2)${}^2$ metrics to raise, lower, and contract indices.  Each component field, as a function of $x$, satisfies the Klein-Gordon equation, and each non-scalar satisifies a Weyl equation (which for $f$ is the combination of the usual field equation and Bianchi identity for the Yang-Mills field strength).  Note that all component fields appear at $y=0$, but some only with $x$ derivatives; as stated above, this is a general feature, following from the equation $»_Ï »_{ÐÏ} +»_x »_y=0$; the same is not true off shell.

The field equations for this case are also implied by the combination of Taylor expandability with the ``reality" condition:
$$ \C ì = ì,ââr+Ðr = 1 $$
where the latter equation implies a factor of $det¼y=y^2$ (4-vector square of the 4 $y$'s) under charge conugation.  (Similar remarks apply for the N=2 scalar multiplet, also described by a scalar with $r+Ðr = 1$.  In that case, since $\C^2 = (-1)^{(N/2)(r+Ðr)}$, the field is pseudoreal, so the field is doubled and satisfies $(\C ì)^i=C^{ij}ì_j$.  In the interacting case, the reality condition for the scalar hypermultiplet involves the prepotential for the vector hypermultiplet.  That prepotential is also real, but it has superscale weight 0, so no field equations are implied for it.)

Ü4.2. Supertwistors

The supertwistor [11] representation is a direct generalization of the bosonic case.  Now we find it as a way to solve the above equations of motion:  Direct substitution of
$$ G_\M{}^\N = ü[н_\M, ½^\NÕ = н_\M ½^\N -ü¶_\M^\N $$
$$ Ó Ð½_\M , ½^\N ] = ¶_\M^\N ,âÓ½,½] = Óн,н] = 0 $$
verifies that it is a solution.  (Just commute the $½$'s together so the symmetrization gives commutators.  Remember that twistors have statistics opposite to those suggested by the indices.  We have used a symmetric ordering as the definition of normal ordering in this case, as in the analogous case of Dirac $©$-matrices, for hermiticity.  However, this is again ambiguous because of the Abelian gauge invariance.)  Note that the supertwistor representation is also a projective space:  Besides dividing up the N-valued $Ða$ index as n + (N$-$n) for arbitrary n, we could have done the same for 4-valued index $Ќ$.  (This would do the same kind of thing for the $x$ coordinates as we have done for the $y$'s.)  The 0+4 case is trivial (it gives no spacetime coordinates), the 2+2 case gives normal 4D spacetime as discussed above, while the 1+3 case gives supertwistors.  However, this would give a complex space, so we need to include the complex-conjugate twistor to define real fields.  Identifying the complex conjugate with the canonical conjugate (as for creation and annihilation operators) then prevents doubling the dimension of the space.  

The supertrace piece $str¼G ­ (-1)^\M G_\M{}^\M$ commutes with the superconformal generators, and should not be considered part of the superconformal group:  It's the superhelicity.  For the ÓoffÕ-shell representation of the previous section,
$$ superhelicity ­ str¼G = str¼D = str¼s_u - str¼s_{Ðu} $$
The superhelicity is part of the ``spin", and becomes nontrivial when thus relaxing the ``S" constraint of the superconformal group, which we treated as a gauge condition.  It's related to the Abelian gauge invariance $¶G¾I$ we considered, except in the case N=4, where $str¼I=0$, and the latter gauge invariance is the definition of the ``P" in ``PSU(4|2,2)".  In the twistor representation, it counts the number of $н$'s minus $½$'s.

This choice of $G$ satisfies exactly
$$ G_{(\M}{}^{(\P}G_{\N]}{}^{\Q]} = ü ¶_{(\M}^\P ¶_{\N]}^\Q $$
With a particular choice of Abelian gauge parameter, the projective superspace representation of the previous section also solves it on shell (without spin).  With another choice of this parameter, one can instead obtain either of
$$ G_{(\M}{}^{(\P}G_{\N]}{}^{\Q]} = à ¶_{(\M}^{(\P} G_{\N]}{}^{\Q]} $$
for both cases.  (Note however that such redefinitions change the relation of $str¼G$ to the superhelicity.)

The super Penrose transform then gives the solution to the equations of motion in projective superspace by identifying $G_w$ in the two representations:  For scalars,
$$ -i»_{M'}{}^M = àн_{M'} ½^MâÜâ
	ì(w) =  Ý_à Çd½Êdн¼e^{ài½wн}_à(½,н) $$
(restoring the ``$i$" for hermiticity), relating the projective superfield $ì(w_M{}^{M'})$ with the twistor superfields $_à(½^M,н_{M'})$ for positive and negative-energy solutions.  The choice of n determines how the fermionic twistor coordinates are distributed between $½$ and $н$.  Note that, unlike $½^\M$ or $н_\M$, these coordinates are not a representation (but only a nonlinear realization) of the superconformal group:  For example, the conformal boosts are represented as quadratic in their ``momenta".

The usual component (bosonic-)twistor fields are obtained by evaluating the expansion of $$ over the fermionic $½$'s.  The expansion in $y$ gives new component fields, but the expansion terminates because of the anticommutativity of the corresponding $½$'s.  The expansion in $Ï$ (and $ÐÏ$) also gives new component fields, but with spinor indices from bosonic $½$, which then satisfy the usual Weyl equation (as in the nonsupersymmetric twistor formalism), and faster termination because there are fewer fermionic $½$'s than $Ï$'s, and because $y$ dependence may give extra fermionic $½$'s.  Also note that expansion in both $Ï$ and $ÐÏ$ will give both $½^µ$ and $н_{Àµ}$, which is equivalent to an $x$ derivative.  We also see that all fields with $y$ dependence also occur without $y$, but with $x$ derivatives, because fermionic $½$'s can come from either $Ï$ or $y$ (but $y$'s give only equal numbers of $½^µ$ and $н_{Àµ}$).  All of this agrees with our previous evaluation in terms of the field equations directly.

As usual, the twistor superfields can be Fourier transformed to functions of just $½^\M$ (or just $н_\M$, or something in-between):  Integrating over just $н_{M'}$,
$$ ì(w) =  Ý_à Çd½^M¼÷_à(½^M,-½^N w_N{}^{M'}) 
	= Ý_à Çd½^\M¼¶(½^{M'}+½^N w_N{}^{M'})¼÷_à(½^\M) $$
In the last form, or the analog from integrating out $½^M$ instead, the argument of the $¶$ function can be replaced with
$$ ½^\M Ðz_\M{}^{A'}âorâz_A{}^\M н_\M $$
since the $u$ dependence factors out as a Jacobian $sdet$.  (If we keep this factor, we get $\on\circ ì$ instead of $ì$, with the correct physical value of $r+Ðr=1$.)

Introducing spin, we find that already the spin-dependent equations appearing in the first set of reduced equations of the previous subsection (i.e., those that also contain derivatives) restrict the supertwistor space solutions to the analog of those for the bosonic case:
$$ ì_{M'...N'}{}^{M...N} (w) =  Ý_à Çd½Êdн¼e^{ài½wн}¼Ð½_{M'}òн_{N'}½^Mò½^N _à(½,н) $$
Since a $½$ and $н$ are produced by a $w$ derivative, this effectively reduces $ì$ to have only unprimed or only primed indices, graded antisymmetric in all of them, as implied by the second (spin-only) set of superconformal field equations.  (However, fields that are total derivatives on shell need not be so off; but such field strengths are generally not conformal.)  In the purely $½^{\M}$ or $н_{\M}$ form, the full indices can be used, but because of the constraint enforced by the $¶$ function, the fields will satisfy the analogous constraints on the indices, as described in the previous section.  The superhelicity is now given by the number of unprimed minus primed indices.

Ü4.3. Propagators

These on-shell properties are enough to write a propagator:  Using the Penrose transform, written as
$$ \on\circ ì = Çd½^\M¼¶(½^\M Ðz_\M{}^{A'}) (½^\M) $$
we can write a propagator as a sum over physical states, Penrose transformed to the endpoints, as
$$ ë = Çd½^\M¼¶(½^\M Ðz_{1\M}{}^{A'}) ¶(½^\M Ðz_{2\M}{}^{A'}) $$
Using the orthogonality of $z$ and $Ðz$, we can solve either $¶$ function as
$$ ½^\M = ½^A z_A{}^\M $$
for some $½^A$, giving
$$ ë = Çd½^A¼¶(½^A z_{1A}{}^\M Ðz_{2\M}{}^{A'}) 
	= sdet(z_{1A}{}^\M Ðz_{2\M}{}^{A'}) $$
effectively from the Jacobian of the $¶$ function.

A nontrivial example is the chiral-antichiral propagator for the N=1 scalar multiplet.  Note that in this case we use 2 different projective superspaces: chiral for $z_1$, antichiral for $Ðz_2$.  Using the usual (in the gauge $u=Ðu=1$)
$$ z_{1Œ}{}^\M = (-Ï_{1Œ},¶_Œ^µ,-x_{1Œ}{}^{Àµ}),ââ
	Ðz_{2\M}{}^{ÀŒ} = (ÐÏ_2^{ÀŒ},x_{2µ}{}^{ÀŒ},¶_{Àµ}{}^{ÀŒ}) $$
we have
$$ ë = {1\over (x_{12}+Ï_1 ÐÏ_2)^2} $$
which agrees with the usual result before detaching the $Ðd^2$ and $d^2$ at the ends,
$$ ë = Ðd_1^2 d_2^2 ¶^4(Ï_{12}){1\over x_{12}^2}  $$
in the chiral representation for 1 and antichiral for 2.

For (real) projective superspace, the propagator is simply
$$ ë = sdet(w_{12}) $$
(We have gauged away any $sdetÊu$ factors:  This is a $ìì$ propagator, not $\on\circ ì\on\circ ì$.)  This includes N=0, where $ë=1/x_{12}^2$, as well as the N=2 result for the scalar hyermultiplet.  The N=4 case gives the propagator for the field strength of the vector multiplet.  It can also be analyzed by components:  Expanding the explicit expression
$$ sdet(w) = {(y-Ïx^{-1}ÐÏ)^2\over x^2} $$
in $Ï$ and $y$ (corresponding to expansion of the associated field strengths) shows the usual propagators for the scalars ($1/x^2$) and spinors ($x/(x^2)^2$), and the field strengths for the vectors ($Òf ÐfÔ=xx/(x^2)^3$), and derivatives of these fields.

These propagators are a bit of a fudge:  They are really ``cut" propagators, homogeneous solutions to the wave equations obtained by summing over physical (positive energy), on-shell states.  However, the St¬uckelberg-Feynman propagator can be obtained by taking this propagator for positive energy and using it for positive times (multiplying by $Î(x_{12}^0)$) and adding it to the negative-energy propagator for negative times.  More simply, one can just fix the $i·$ prescription by hand:  For example, for N=0 we can write the twistor integral as (in half-Fourier-transformed twistor variables)
$$ Çd½^ŒÊdн^{ÀŒ}Êe^{ài½^Œ н^{ÀŒ} x_{ŒÀŒ}} = {1\over x^2} $$
To make this converge, we need 
$$ x_{ŒÀŒ} £ x_{ŒÀŒ} ài·¶_{ŒÀŒ} $$
Since the identity part of the matrix $x_{ŒÀŒ}$ corresponds to the time component $x_0$, as in $e^{ài|p^0|x_0}$, this implies
$$ {1\over x^2} £ {1\over x^2 ài·x^0} $$

\noindent with signs corresponding to those in the integral.  By comparison, the Feynman propagator is $1/(x^2+i·)$.

Û6 5. AdS/CFT

Ü5.1. Projective lightcone limit

The Anti de Sitter/Conformal Field Theory correspondence proposes to relate 4D N=4 (super)conformal field theory to IIB superstring theory expanded about the 10D (bosonic) manifold of 5D anti de Sitter space $ð$ the 5-sphere.  This incorporates the use of ``holography":  Instead of the usual procedure of solving the wave equation for time dependence, we solve it for spatial dependence, on the coordinate whose endpoint defines the boundary of AdS$_5$.

As a simple, well-known example we consider a free, massless scalar field on AdS$_{D+1}$.  
In Poincar«e coordinates, the metric is
$$ -ds^2 = R^2{dx^2 +dx_0^2\over x_0^2} $$
in terms of the AdS radius $R$, where $x_0$ is a spatial coordinate.  A simple description is obtained by embedding in flat space, as
$$ X^2 = -R^2âÜâX = (X^+, X^i, X^-) = R{(1, x^i, ü(x^2 + x_0^2))\over x_0} $$
with $dX^2=ds^2$ as above.  (We use the lightcone basis $X^2=(X^i)^2-2X^+ X^-$.)  This description is equivalent to that in terms of the coset SO(D,2)/SO(D,1); however, it's clearly simpler to start with a (D+2)-component vector than a (D+2)(D+1)/2-component matrix.

The (singular) boundary limit $x_0£0$ is equivalent to the limit $R£0$, describing flat space of one less dimension [12].  
This can be interpreted as the relation between active and passive approaches:  Instead of (Muhammad) moving to the boundary, we shrink the distance scale, effectively moving the boundary closer.  (It is a type of long-distance limit, in contrast to the short-distance limit $R£¥$ related to flat space.)  In terms of the above metric, we first rescale
$$ x_0 £ R x_0âÜâ
	-ds^2 £ {dx^2 +R^2 dx_0^2\over x_0^2} $$
(Alternatively, we can scale $x£x/R$ instead.)  The limit $R£0$ pinches AdS into a lightcone, reducing the conformal analysis to that of the projective lightcone. 

This limit contracts the gauge group SO(D,1) of the coset to ISO(D$-$1,1), while leaving the symmetry group SO(D,2) intact.  
In this limit $x_0$ survives in a trivial way:  It makes $dx^2/x_0^2$ conformally invariant, and gives a simple way of seeing it.  For purposes of describing just this flat space, it can be removed by introducing a ``projective" scale invariance $¶X=Â(X)X$.  
The only symmetry generators with dependence on $x_0$ are the dilatation and conformal boosts.  Explicitly, we have
$$ ë = xÉ»_x + x_0 »_{x0} $$
$$ K_a = x_a(xÉ»_x + x_0 »_{x0}) -ü(x^2 +R^2 x_0^2)»_a $$
(In this subsection ``$ë$" has nothing to do with propagators.)  In the limit $R£0$, these take the D-dimensional flat-space form, with $x_0 »_{x0}$ acting as the scale weight.

In this limit the free, massless scalar field becomes
$$ \on\circ Ä(x,x_0) £ x_0^{(D-2)/2}Ä(x) $$
replacing the original solution of the AdS Klein-Gordon equation with a solution of that in Minkowski space of one less dimension.  Again $x_0$ appears trivially, making the field a scalar under conformal transformations.  It also follows directly from the projective ($R=0$) formulation, as a result of the projective invariance $XÉP+PÉX=0$ that follows from the closure of $X^2=0$ and $P^2=0$.
In this case
$$ D=4âÜâ\on\circ Ä(x,x_0) £ x_0 Ä(x) $$

This result also follows from dimensional analysis:  Since the original field $\on\circ Ä$ was a ``scalar" under the conformal group (i.e., had vanishing scale weight), the engineering dimension of the usual field $Ä(x)$ must be canceled by an appropriate power of $x_0$.  In the usual holographic analysis, this is identified with holographic-``time" dependence:  If we write $x_0=e^{-t}$, so the corresponding term in the metric is simply $dt^2$, then the dependence of a field in the limit $t£¥$ ($x_0£0$) is $e^{-t÷ë}$, where $÷ë$ is the conformal weight.  Amputation of this factor in AdS amplitudes is then equivalent to use of the interaction picture for this Euclidean time coordinate.

Ü5.2. Superlimit

Here we'll apply a slightly different procedure:  In lightcone quantization the wave equation is solved for dependence on a lightlike coordinate.  Furthermore, for applying twistor techniques to Feynman diagrams it's convenient to Wick rotate this idea to ``spacecone" quantization, using a complex, null, spatial coordinate [13].  We'll find it convenient to use a similar procedure here, to find the correspondence between the superspaces of AdS and CFT.

To see why such a treatment naturally arises, we work in Poincar«e coordinates for S$^5$, after an appropriate Wick rotation.  Combining the two spaces (with the signs that follow from the grading),
$$ ds^2 = {dy^2 + R^2 dy_0^2\over y_0^2} -{dx^2 + R^2 dx_0^2\over x_0^2}  = 
{dy^2\over y_0^2} - {dx^2\over x_0^2} + R^2Êd¼ln(x_0 y_0) d¼ln(y_0/x_0) $$
We can then identify $x_0 y_0$ and $y_0/x_0$ (or some functions of just one or just the other) as two null, spatial coordinates, to be used to define our spacecone quantization.  

We then modify the usual boundary limit of AdS to $x_0 y_0£0$ ($y_0/x_0$ fixed), in line with interpretation of $x_0 y_0$ as the spacecone ``time".  This leaves us with 9 bosonic coordinates on the boundary, 8 of which have translation invariance, and are to be identified with the 4 $x$'s and 4 $y$'s of 4D N=4 projective superspace.    (There is a symmetry under translation of the 9th coordinate, but it requires also scaling of the other 8, as well as the fermions.  It is associated with a combination of a dilatation with an R-symmetry U(1).)

To generalize this limit to superspace, and see how it naturally arises in the projective approach, consider a general supergroup element of PSU(4|2,2), which is a symmetry on both the AdS and CFT sides (hence the correspondence).  We want to define the boundary limit as one which picks out N=4 projective superspace, while preserving this symmetry (but perhaps not the gauge groups).  Knowing how the projective space fits into the group element (and its inverse), this limit must be the $R£0$ limit after the rescaling
$$ g_\M{}^\A £ \left(åR g_\M{}^A , {1\over åR}Ðz_\M{}^{A'}\right),ââ
	g_\A{}^\M £ \left({1\over åR}z_A{}^\M , åR g_{A'}{}^\M \right)  $$
Note that the scaling by $R$ is determined only by the $\A$ index, and is independent of the symmetry index $\M$.

  This limit eliminates the $v$ coordinates, which don't appear in the projective approach, and leaves $w$, but also some parts of $u$ and $Ðu$, depending on the choice of gauge group.  After eliminating $v$, and expressing $z$ and $Ðz$ in terms of the rest, we see the $R$ scaling is
$$ w £ w,ââu £ åRu,ââÐu £ åRÐu $$

In particular, it's easy to pick out $x_0$ and $y_0$ as the pieces of $u$ and $Ðu$ invariant under the manifest SO(3,1) Lorentz and SO(4) internal symmetries, after killing the ``PS" pieces of PSU(4|2,2):
$$ u = \pmatrix{ å{y_0}ÊI & 0 \cr 0 & å{x_0}ÊI \cr}u_0,ââ
	Ðu = \pmatrix{ å{y_0}ÊI & 0 \cr 0 & å{x_0}ÊI \cr}Ðu_0 $$
$$ sdet¼u_0 = sdet¼Ðu_0 = det¼u_0 = det¼Ðu_0 = 1 $$
This can be seen, e.g., by considering the N=0 case, and noting that there $det(zdÐz) =dx^2/x_0^2$ is the metric of the projective lightcone.  We then have 
$$ sdet¼u = sdet¼Ðu = {y_0\over x_0} $$
Then we see that the $R£0$ limit is the boundary limit, since now the scaling is on just $y_0£Ry_0$ and $x_0£Rx_0$:
$$ w £ w,ââu_0 £ u_0,ââÐu_0 £ Ðu_0,ââ{y_0\over x_0} £ {y_0\over x_0};ââ
	x_0 y_0 £ R^2 x_0 y_0 $$

Effectively we have defined the coordinate $x_0 y_0$ by including it in exactly (and only) the same way $R^2$ was introduced in the rescaling.

Ü5.3. 10D field equations

The correspondence relates fundamental fields in the string theory to color-singlet composite fields in the conformal field theory.  Of particular interest are fields that correspond to reducing the superstring to a superparticle:  They describe 10D IIB supergravity (again perturbed about AdS$_5ð$S$^5$).

Both 10D IIB supergravity and 4D N=4 super Yang-Mills are representations of the group PSU(4|2,2).  But the physical interpretation is different:  For example, they satisfy different field equations, even at the free level.  We saw the free field equations for (the field strengths of) 4D super Yang-Mills, and applied them in projective superspace.  On the other hand, 10D supergravity satisfies different, weaker equations (since more dimensions):  Its free field equations are [14]
$$ G_\M{}^\P G_\P{}^\N = 0¼mod¼¶¼terms $$
In the boundary limit these are not the 4D Yang-Mills equations, but the equations satisfied by BPS color-singlet composites of the Yang-Mills superfields.  

We saw the stronger equations implied $p^2=0$ in D=4 by picking indices giving the highest (engineering) dimension; thus the rest of the equations followed by conformal supersymmetrization.  That was easy, since all 4 indices were free in that case, whereas here some are contracted.  Now we restrict to the bosonic sector of the weaker 10D equations, which is sufficient, as the supersymmetric generalization is unique.  This means we truncate the symmetry group to SU(4)$°$SU(2,2), which is not the same as considering the N=0 case.  The field equations are then of the form
$$ G_{Ðm}{}^{Ðp}G_{Ðp}{}^{Ðn} = ¶_{Ðm}^{Ðp} \O,ââG_{е}{}^{Ш}G_{Ш}{}^{ÐÃ} = ¶_{е}^{ÐÃ}\O $$
for some operator $\O$.  These can be translated into vector notation as
$$ G_{[\underline{mn}}G_{\underline{pq}]} = G_{[\underline{µÃ}}G_{\underline{¨§}]} = 
	G^{\underline{mn}}G_{\underline{mn}} - G^{\underline{µÃ}}G_{\underline{µÃ}} = 0 $$
which generalize to arbitrary AdS$_mð$S$^n$, where $\un m$ and $\un µ$ are vector indices for SO(n+1) and SO(m-1,2).  If we plug in the usual representations of these symmetry groups on these spaces, then the former 2 equations say that the corresponding spins vanish, while the last is the Klein-Gordon equation in m+n dimensions.   If we had set $\O$ to vanish, decoupling the 2 spaces, we would instead have the m-dimensional Klein-Gordon equation on AdS, while on the sphere we would leave only a constant solution.

In fact, the supersymmetric 10D equations above are satisfied by Óoff-shellÕ 4D N=4 projective superpspace (as seen after converting to the $DD$ form of the equations, by multiplying by $g$ and $g^{-1}$ appropriately to convert indices):  As a consequence of reducing to just the $w$ coordinates, the $D$'s have only lower primed indices and upper unprimed, which can't be contracted.  This result can be generalized a bit:  
\vskip-.1in
$$ I\O ¾ \pmatrix{ D_u & -D_v \cr D_w & -D_{Ðu} \cr}^2 = 
	\pmatrix{ D_u^2 -D_v D_w & -D_u D_v +D_v D_{Ðu} \cr
		D_w D_u -D_{Ðu}D_w & -D_w D_v +D_{Ðu}^2 \cr} $$
		
\noindent Still applying $D_v=0$ (to allow the projective approach) and leaving $D_w$ unconstrained, we can consider modifying the $D_u$ and $D_{Ðu}$ constraints, as we did when considering arbitrary (super)spin.  
The solution is that the field is a scalar, which we already knew was true by construction as a superparticle, since no spin degrees of freedom were introduced.  More specifically, we find
$$ D_A{}^B = -r¶_A^B,ââD_{A'}{}^{B'} = -Ðr¶_{A'}^{B'};ââr = Ðr $$
for some ``central charge" $r$ that commutes with $D_w$.  From the previously given solution to this constraint for any eigenvalue of $r+Ðr$,
$$ \on\circ ï(w,u,Ðu) = (sdet¼u)^{-r}(sdet¼Ðu)^{-Ðr}ï(w) $$
(we should solve before setting $sdet¼u=sdet¼Ðu$, etc.) and our above choice for defining $y_0/x_0$, we see that
$$ r = Ðr = - {»\over »¼ln(y_0/x_0)} $$
and our general solution to the 10D field equations is in terms of a field that is an arbitrary function of $w$ and $y_0/x_0$.  (The same result is obtained if we calculate directly $D_u=»_u u$, etc., paying careful attention to signs from the grading.  For example, $»_M{}^A u_B{}^N = ¶_A^B ¶_M^N$ has an implicit factor of $(-1)^A$ from the $A$ being to the left of the $B$.)  

Thus, in the same way that 4D supertwistors solve the free 4D field equations, 4D N=4 projective superspace (plus the coordinate $ln(y_0/x_0)$) can be considered to be the supertwistor space of free 10D IIB supergravity on AdS$_5ð$S$^5$.  It solves these 10D field equations in terms of ``initial conditions" (in the spacecone sense) at the 9D boundary $x_0 y_0 =0$.

Ü5.4. Correspondence

We now investigate the significance of this 9th coordinate $x_0/y_0$ to the CFT.  Consider expansion of the 10D theory over S$^5$ in terms of spherical harmonics.  These can all be expressed in terms of those for the vector harmonic, which are given by a unit 6-vector; in the coordinates we've been using, these are
$$ Y = (Y^+, Y^i, Y^-) = {(1, y^i, ü(y^2 + y_0^2))\over y_0} $$
In the boundary limit, this becomes a null 6-vector,
$$ Y £ {(1, y^i, üy^2)\over y_0} $$
homogeneous in $y_0$.  This $y$ dependence can clearly be associated with that of the scalars of 4D N=4 Yang-Mills, i.e., the field strength $ì$ at $Ï=0$.
A similar analysis can be made for the $x_0$ dependence of the scalars, as discussed above.  (In general, interactions modify this result; but for the fundamental fields of 4D N=4 Yang-Mills, and the BPS composite operators considered here, ultraviolet finiteness preserves conformal weights.)  

We can easily supersymmetrize this result to identify the other fields of the supermultiplet, and see how they appear in color singlets.  Returning to our analysis of general spin, noting that $ì$ is a scalar with $r+Ðr=1$, and again substituting for the $sdet$, we have
$$ \on\circ ì(w,u,Ðu) = {x_0\over y_0}Êì(w) $$
reproducing the $x_0$ and $y_0$ dependence found above for the scalars.  ($x_0$ dependence is determined by the superscale weight of the multiplet, and $y_0$ by the super-U(1) weight.  The corresponding symmetry generators also have $Ï»_Ï$ terms, giving different component scale and U(1) weights to the higher spins.)  It then follows that the supergravity superfield source on the boundary must take the form
$$ \on\circ ï\left(w,{x_0\over y_0}\right) = 
	tr\leftÓf\left[\on\circ ì\left(w,{x_0\over y_0}\right)\right]\rightÕ $$

\noindent for some (Taylor expandable) function $f$, and thus contains terms of the form
$$ tr \leftÓ \left[ {x_0\over y_0}Êì(w) \right]^n \rightÕ $$

\noindent Thus, the 9th bosonic coordinate on the boundary just counts the number of supergluons.  Note that, unlike the usual $x_0£0$ limit, in this limit the supergravity fields are nonvanishing, having no dependence on $x_0 y_0$ (but string excitations will have positive powers of $x_0 y_0$, corresponding to anomalous dimensions in the 4D field theory).  Also for these supergravity fields on the boundary, the ``momentum" conjugate to the coordinate $ln(x_0/y_0)$ is quantized.

The 10D supergravity superfield is real.  ($Y$ is real:  Because of Wick rotation, $Y^i$ is real but $Y^+*=-Y^-$.  This implies the usual charge conjugation for $y$ on the boundary, while $y_0$ gives the density part of charge conjugation to $ì$.)  It's also nonsingular on S$^5$:  Since it can be expanded in spherical harmonics, that means on the boundary only nonnegative powers of $y$ will appear.  Thus $ì$ is forced to satisfy its (ÓinteractingÕ) field equations.

Ü5.5. Bosonic propagator

The bosonic scalar propagator can be easily derived by direct solution of the Klein-Gordon equation.  We give the calculation here as an illustration of some of the properties of spacecone quantization.  The metric in spacecone coordinates is
$$ ds^2 = {e^{-x^-}dy^2 - e^{x^-}dx^2 +dx^+ dx^-\over x^+} $$
where
$$ x^+ = x_0 y_0,âââx^- = lnÊ{y_0\over x_0} $$
have been chosen so that the metric is homogeneous in $x^+$ and symmetric in $x$ and $y$, up to signs.  ($x^+$ is then the natural worldvolume coordinate ``$ $" in first-quantization schemes.)

The Klein-Gordon equation on AdS$_{D+1}ð$S$^{D+1}$ for the propagator
$$ {1\over å{|g|}}»_m å{|g|}g^{mn}»_n ë ¾ {1\over å{|g|}}¶^{2(D+1)} $$
factoring out the $1/å{|g|}=(x^+)^{D+1}$, is then given by the operator
$$ (x^+)^{-D}\left(e^{x^-}»_y^2 -e^{-x^-}»_x^2 +»_+ »_- -{D\over 2x^+}\right) $$
$$ = (x^+)^{-D/2}(e^{x^-}»_y^2 -e^{-x^-}»_x^2 +»_+ »_-) (x^+)^{-D/2} $$
We then move the $x^+$ factor on the left to the right-hand-side of the equation to convert it to an $(x^+)'$, then move it back again, to get
$$ (e^{x^-}»_y^2 -e^{-x^-}»_x^2 +»_+ »_-) ÷ë ¾ ¶^{2(D+1)},ââ÷ë = (x^+ x'^+)^{-D/2}ë $$
We thus have a nonrelativistic-style Schr¬odinger equation of the form $(p^--H)Æ=0$ with $x^+$-independent ``Hamiltonian" $H$.

At this point the equation is separable in not only $y$ and $x$, but also $x^+$, so we can Fourier transform in those coordinates to get a Ófirst-orderÕ differential equation (``time-independent Schr¬odinger equation") in just $x^-$.  The solution is of the form
$$ ÷ë ¾ 
   àÎ[à(x^--x'^-)]Ê{1\over p_+}Êexp\left(iÊ{e^{x^-}p_y^2 +e^{-x^-}p_x^2\over p_+}\right) $$
plus a term independent of $x^-$ (except for the $Î$), which has $x'^-$ instead.  The Heaviside step function $Î$ turns the homogeneous solution into an inhomogeneous one; the choice of $à$ will be discussed shortly.  We next Fourier transform back in $y$ and $x$ to put $p_+$ in the numerator of the exponent, so the final Fourier transform in $x^+$ will be trivial.  (It also gives a measure factor $(p_+)^D$.)

We then use the $i·$ prescription that positive $p^-$ ($¾p_+$) propagates forward in $x^-$ and negative backwards.  (Normally this is applied to the time $x^0$ or lightcone ``time" $x^+$, but it can be applied to any coordinate.  This follows from first-quantization or Schwinger parametrization as $dx^m=d ¼p^m$ with $d $ positive.)  The integral over  this sign-restricted $p_+$ then gives the known result
$$ ë ¾ (s_y-s_x-i·)^{-D},ââs_y -s_x = {(y-y')^2 +(y_0-y'_0)^2\over 2y_0 y'_0} 
	- {(x-x')^2 +(x_0-x'_0)^2\over 2x_0 x'_0} $$

Ü5.6. String cosets

So far we have avoided specifying the precise superspace used for describing the superstring.  However, we have seen how 4D N=4 projective superspace can arise from taking the appropriate boundary limit, and how it appears upon solving the 10D equations of motion for the superparticle (supergravity).  Certainly the string superspace must include at least these coordinates (and $x_0 y_0$).  The action must also explicitly depend on $R$, as defined above in terms of the group elements.  (For example, we saw for N=0 that $R$ appears in the metric for AdS coset, but drops out in the flat-space coset.)

Superstrings, like superparticles, can be quantized in spacecone gauges.  In such gauges (like lightcone gauges) only the physical fermions survive, 1/4 of the fermions of the full superspace.  Also, all the physical bosons survive, corresponding to the dimension of spacetime.  However, in the string case there are oscillators, associated with excitations to massive levels of the string, only for the transverse dimensions.  Thus here we expect 8 fermions (and their canonical conjugates), with their associated oscillators, and 10 bosons, only 8 of which have oscillators.  The 8 fermions + 8 bosons are associated directly with the 4D N=4 projective coordinates, while the 2 remaining ``zero-modes" are associated with $x_0$ and $y_0$.  Thus the AdS/CFT correspondence for the superspaces is clear in the spacecone gauge.

We also know that the 10D superspace in a general gauge must have more fermionic coordinates than the 8 fermions of the spacecone-gauge superspace:  In particular, the full 32 fermions of the group PSU(4|2,2) correspond directly to those of the full superspace of 10D IIB supergravity.  The full superspace is then the coset PSU(4|2,2)/USp(4)$°$USp(2,2) [15].  However, it might also be useful to employ an intermediate superspace with 16 fermions, analogous to the projective and (anti)chiral superspaces of D=4.  Chiral and antichiral 10D IIB superspaces for supergravity are straightforward:  Under the gauge group USp(4)$°$USp(2,2), the 32 fermions divide up into a complex 4$ð$4 and its complex conjugate.  Chiral superspace uses just one of these, antichiral just the other.  So either of these subspaces can be written as PSU(4|2,2)/I[USp(4)$°$USp(2,2)], where the ``I" refers to inhomogeneous.  This preserves the symmetry because the complex fermion (which was a 16-component 10D spinor in flat space) has a charge under a U(1) that isn't part of the symmetry algebra, and its complex conjugate the opposite charge, while the bosons are all neutral (as for 4D N=1).  As a result, its covariant derivatives anticommute with themselves, and so can consistently vanish.  Unfortunately, this U(1) symmetry of the superparticle action (describing supergravity) is not a symmetry of the superstring action (describing also massive 10D fields), so chiral superspace is not defined for the whole superstring.  

A possible alternative is a projective-like superspace PSU(4|2,2)/OSp(4|4), which picks a real combination of the fermions, but leaves more bosons in the ``internal" space, SU(4)/SO(4).  These might be interpreted as the internal space accompanying the 10D spacetime, just as 4D projective superspace needs $y$ coordinates in addition to the 4 spacetime coordinates $x$.  This space follows from the full superspace by (1)¼applying a first-class subset of the fermionic constraints (still as linear constraints), and (2) relaxing some of the internal USp(4) constraints, for separate application (perhaps in quadratic form).

ÜAcknowledgments

I thank Machiko Hatsuda and Luca Mazzucato for helpful discussions.  This work is supported in part by National Science Foundation Grant No.¼PHY-0653342.

\refs

£1 
  L. Andrianopoli and S. Ferrara,
  ÓPhys. Lett.  BÕ {\bf 430} (1998) 248
  \xxxlink{hep-th/9803171};
\\
  H. Ooguri, J. Rahmfeld, H. Robins and J. Tannenhauser,
  ÓJHEPÕ {\bf 0007} (2000) 045
  \xxxlink{hep-th/0007104};
\\
  P. Heslop and P.S. Howe,
  ÓPhys. Lett.  BÕ {\bf 502} (2001) 259
  \xxxlink{hep-th/0008047}.

£2 
  H. Dorn, M. Salizzoni and C. Sieg,
  ÓJHEPÕ {\bf 0502} (2005) 047
  \xxxlink{hep-th/0307229};
\\
  P. Dai, R. N. Huang and W. Siegel,
  ÓJHEPÕ {\bf 1003} (2010) 001
  \xxxlink{0911.2211} [hep-th].

£3 
  A. Galperin, E. Ivanov, S. Kalitsyn, V. Ogievetsky and E. Sokatchev,
  ÓClass.\ Quant.\ Grav.\  Õ{\bf 1} (1984) 469;
  \\
  A. Galperin, E. Ivanov, V. Ogievetsky and E. Sokatchev,
  ÓJETP Lett.\  Õ{\bf 40} (1984) 912
  [ÓPisma Zh.\ Eksp.\ Teor.\ Fiz.\  Õ{\bf 40} (1984) 155];
  \\
  B.M. Zupnik,
  ÓTheor.\ Math.\ Phys.\  Õ{\bf 69} (1986) 1101
  [ÓTeor.\ Mat.\ Fiz.\  Õ{\bf 69} (1986) 207];
\\
  G.G. Hartwell and P.S. Howe,
  ÓInt.\ J.\ Mod.\ Phys.\  A Õ{\bf 10} (1995) 3901\\
  \xxxlink{hep-th/9412147},
  ÓClass.\ Quant.\ Grav.\  Õ{\bf 12} (1995) 1823;
  \\
  P. Heslop and P.S. Howe,
  ÓClass.\ Quant.\ Grav.\  Õ{\bf 17} (2000) 3743
  \xxxlink{hep-th/0005135};
  \\
  P.J. Heslop,
  ÓClass.\ Quant.\ Grav.\  Õ{\bf 19} (2002) 303
  \xxxlink{hep-th/0108235}.

£4 
  W. Siegel,
  ÓPhys.\ Rev.\  D Õ{\bf 52} (1995) 1042
  \xxxlink{hep-th/9412011}.

£5 A.S. Galperin, E.A. Ivanov, V.I. Ogievetsky, and E.S. Sokatchev, ÓHarmonic superspaceÕ
(Cambridge Univ. Press, 2001) p. 177, footnote.

£6 
  A. Karlhede, U. Lindstr¬om and M. Ro×cek,
  ÓPhys.\ Lett.\  B Õ{\bf 147} (1984) 297;
  \\
  U. Lindstr¬om and M. Ro×cek,
  ÓCommun.\ Math.\ Phys.\  Õ{\bf 115} (1988) 21,
  {\bf 128} (1990) 191.

£7 
  D. Jain and W. Siegel,
  ÓPhys. Rev.  DÕ {\bf 80}, (2009) 045024
  \xxxlink{0903.3588} [hep-th].

£8 
  M. Hatsuda and W. Siegel,
  ÓPhys. Rev.  DÕ {\bf 67} (2003) 066005
  \xxxlink{hep-th/0211184},
\\
  ÓPhys. Rev.  DÕ {\bf 77} (2008) 065017
  \xxxlink{0709.4605} [hep-th].

£9 W. Siegel, Free field equations for everything, in ÓStrings, Cosmology, Composite Struc- turesÕ, March 11-18, 1987, College Park, Maryland, eds. S.J. Gates, Jr. and R.N. Mohapatra (World Scientific, Singapore, 1987).

£10 A.J. Bracken, ÓLett. Nuo. Cim.Õ É2 (1971) 574;\\
A.J. Bracken and B. Jessup, ÓJ. Math. Phys.Õ É23 (1982) 1925;\\
W. Siegel, ÓNucl. Phys.Õ ÉB263 (1986) 93.

£11 A. Ferber, ÓNucl. Phys.Õ ÉB132 (1978) 55.

£12 P.A.M. Dirac, ÓAnn. Math.Õ É37 (1936) 429;\\
H.A. Kastrup, ÓPhys. Rev.Õ É150 (1966) 1186;\\
G. Mack and A. Salam, ÓAnn. Phys.Õ É53 (1969) 174;\\
S. Adler, ÓPhys. Rev.Õ ÉD6 (1972) 3445;\\
R. Marnelius and B. Nilsson, ÓPhys. Rev.Õ ÉD22 (1980) 830.

£13 
  G. Chalmers and W. Siegel,
  ÓPhys. Rev.  DÕ {\bf 59} (1999) 045013
  \xxxlink{hep-ph/9801220}.

£14 
  M. Hatsuda and K. Kamimura,
  ÓNucl.\ Phys.\  B Õ{\bf 611} (2001) 77
  \xxxlink{hep-th/0106202}.

£15 
  R.R. Metsaev and A.A. Tseytlin,
  ÓNucl. Phys.  BÕ {\bf 533} (1998) 109
  \xxxlink{hep-th/9805028}.

\bye